\documentclass{caps}
\usepackage{graphicx,psfig,latexsym,longtable,lscape,rotating}

\def\kms{\hbox{km s$^{-1}$}}
\def\ergsec{\hbox{erg s$^{-1}$ }}

\def\msun{$M_{\odot}$}
\def\rsun{$R_{\odot}$}

\def\mdot{$\dot M$}

\def\p0{\phantom{0}}

\def\it{\sl}

\DeclareRobustCommand{\ion}[2]{%
\relax\ifmmode
\ifx\testbx\f@series
{\mathbf{#1\,\mathsc{#2}}}\else
{\mathrm{#1\,\mathsc{#2}}}\fi
\else\textup{#1\,{\mdseries\textsc{#2}}}%
\fi}

\makeatletter \renewcommand\@biblabel[1]{$^{#1}$} \makeatother

\begin{document}

\pagenumbering{roman}
\cleardoublepage

\pagenumbering{arabic}

\setcounter{chapter}{4}

\setcounter{table}{0}

\author[P.A. Charles and M.J. Coe]{P.A. CHARLES and M.J. COE\\School of Physics \& Astronomy, University of Southampton,\\ Highfield, Southampton SO17 1BJ, UK}

\chapter{Optical, Ultraviolet and Infrared Observations of X-ray Binaries}


\section{Introduction}

In the 35 years since the first X-ray binary was optically identified
(Sco X-1) the basic division of X-ray binaries into the high-mass
(HMXBs) 
and low-mass (LMXBs) systems has become firmly
established.  The nomenclature refers to the nature of the mass donor,
with HMXBs normally taken to be $\geq$10\msun, and LMXBs $\leq$1\msun.
However, the last decade has seen the identification and measurement
of a significant number of X-ray binaries whose masses are
intermediate between these limits.  Nevertheless, the nature of the
mass-transfer process (stellar wind dominated in HMXBs, Roche lobe
overflow in LMXBs) produces quite different properties in the two
groups and so this chapter will be divided into two main sections on
HMXBs and LMXBs.  A more complete Introduction can be found in chapter 1. 

While the nature of the compact object and its properties are largely
determined from X-ray studies (e.g. chapters 1,4), longer-wavelength
observations allow detailed studies of the properties of the mass
donor.  This is most straight-forward for the intrinsically luminous
early-type companions of HMXBs, which provides the potential for a
full solution of the binary parameters for those systems containing
X-ray pulsars.  This is particularly important for HMXB evolution in
that it allows a comparison of the derived masses with those obtained
for neutron stars in the much older binary radio pulsar systems
(Thorsett \& Chakrabarty, 1999).

However, when HMXBs are suspected of harbouring black holes (e.g. Cyg
X-1), the mass measurement process runs into difficulties.  By
definition, there will be no kinematic features (such as pulsations)
associated with the compact object, and hence the analysis is based
entirely on the mass-losing companion.  Most importantly, the mass of
the compact object depends on the accuracy to which that of the
companion is known.  The situation for LMXBs, such as Sco X-1, is
completely different, in that their short orbital periods require
their companion stars to be of much lower mass.  In
section~{\ref{sect:LMXBs}} we collect together and discuss the results
of mass determinations in all X-ray binaries.  We also emphasise those
properties that LMXBs have in common with cataclysmic variables (chap
10), most of which are related to the similar structure of their
accretion discs.

More recently, the advent of near-IR spectroscopic instrumentation on
large telescopes has allowed the first detailed investigations of
X-ray binaries in the Galactic Bulge and other regions of high
extinction.

\section{\sc High-mass X-ray binaries}

\subsection{Supergiant and Be X-ray binaries}

There are two main sub-groups of HMXBS - the supergiant counterparts
(normally of luminosity class I or II), and the Be/X-ray (or BeX) binary
systems (normally luminosity class III or V). Both sub-groups involve OB
type stars and are commonly found in the galactic plane and the
Magellanic Clouds. They differ, however, in accretion modes with the
supergiant systems accreting from a radially outflowing stellar wind,
and the BeX binaries accreting directly from the circumstellar
disc (possibly with some limited Roche lobe overflow on rare
occasions). As a result the supergiants are persistent sources of
X-rays, whilst the BeX systems are very variable and frequently
much brighter.

Recently Reig \& Roche (1999)  have suggested a third
sub-group of systems; the X Per like systems. Their main
characteristics are: long pulse periods
(typically 1000s), persistent low $L_x$ ($\sim10^{34}$
erg/s) and low variability, and rare uncorrelated weak X-ray
outbursts. The
properties of all three groups are summarised in Table~\ref{hmxb}.

\begin{figure}
\includegraphics[width=3in]{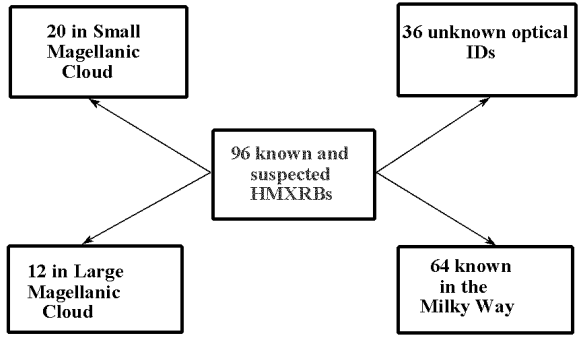}
\caption{Distribution of different HMXB populations}\label{dist}
\end{figure}

Currently (August 2003) there are about 100 known or suspected
HMXBs (see Figure~\ref{dist}). Surprisingly, nearly one third of these
lie in the Magellanic Clouds. This very large fraction, particularily
noticeable in the SMC, will be discussed later. About one third of all
the HMXBs have no known optical counterpart, simply due to the historical inaccuracy of the X-ray
location. Current facilities such as the {\it Chandra X-ray
Observatory} are very effective tools to rectify this situation.


Progress towards a better understanding of the physics of these
systems depends on multi-wavelength studies.  From observations of the
Be star in the optical, UV and IR, the physical conditions under which
the neutron star is accreting can be determined.  In combination with
hard X-ray timing observations, this yields a near complete picture of
the accretion process.  It is thus vital to identify the optical
counterparts and obtain UV to IR measurements for these X-ray systems
in order to further our understanding.

\begin{table}
\caption{General properties of HMXBs. }\label{hmxb}
\begin{tabular}{ccccccc}
\hline
Type$^*$&Percent&Optical   &Typical  &Typical  &Typical     &Log $L_X$\\
&of all &Luminosity&pulse    &binary   &binary      & \\
&HMXB   &Class     &period(s)&period(d)&eccentricity&(\ergsec)  \\
\hline
\hline
Be	&57	  &III-V     &0.05-500 &2-260     &0.3-0.9 &36-38    \\
SG      &25       &I-II      &200-700  &3-40      &        &34-35    \\
XP      &10       &III-V     &200-1400 &250       &0.03    &34-35    \\
Others  &8        & -        & -       &    -     &   -    &    \\
\hline
\end{tabular}
{\footnotesize $^*$ Be = Be star binaries, SG =
supergiant sytems, XP = systems similar to X Per}
\end{table}

\subsection{Be/X-ray binaries}

The BeX binary systems represent the largest sub-class of HMXBs.  Of
96 currently proposed HMXB pulsars, 57\% are identified as BeX type.
The orbit of the Be or supergiant star and the compact object,
presumably a neutron star, is generally wide and eccentric.  X-ray
outbursts are normally associated with the passage of the neutron star
through the circumstellar disc.  The physics of accretion-powered
pulsars has been reviewed previously (e.g. White, Nagase \& Parmar
1995, Nagase 1989, Bildsten et al 1997).  This section will
concentrate on the optical, UV and IR observational properties of BeX
systems and address how these have contributed to our understanding of
Be star behaviour.

X-ray behavioural features of BeX systems include:
\begin{itemize}
\item regular periodic outbursts at periastron ({\it Type I});
\item giant outbursts at any phase probably arising from a dramatic
expansion of the circumstellar disc ({\it Type II});
\item ``missed'' outbursts frequently related to low H$\alpha$
emission levels (hence a small disc), or other unknown reasons (e.g.
perhaps centrifugal inhibition of accretion, see Stella, White \& Rosner
1986);
\item shifting outburst phases due to the rotation of density
structures in the circumstellar disc(Wilson et al, 2002).
\end{itemize}

For supergiant systems the X-ray characteristics tend to be much less
dramatic. Because of the predominantly circular orbits and the much
steadier wind outflow, the X-ray emission tends to be a regular
low-level effect. Rarely Type II outbursts may occur, but never Type
I.

\begin{figure}
\includegraphics[width=3.5in, angle=-90]{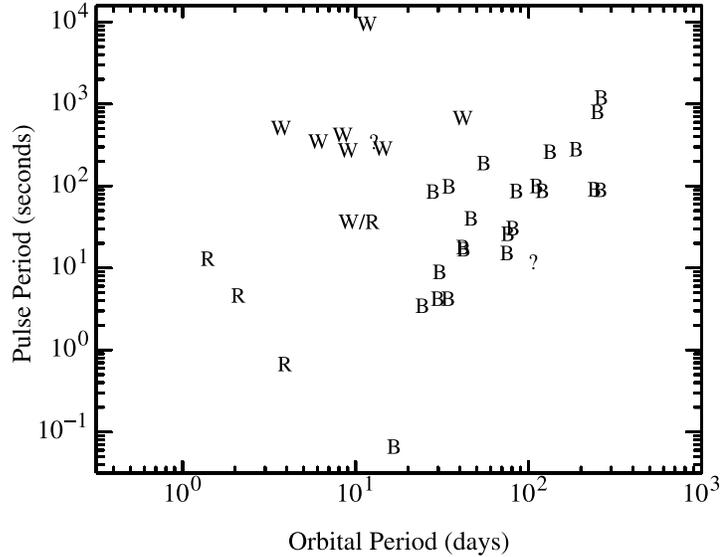}
\caption{An updated version of the Corbet diagram (Corbet 1986) provided
by Corbet (private communication). The
three classes of object illustrated are: B = BeX binaries, W =
wind-fed, R = Roche-lobe overflow. Two systems are
indicated by ? symbols because their optical 
properties are not yet clear.}\label{corbet}
\end{figure}

One particularily interesting piece of astrophysics emerged early on
in the study of BeX systems - the Corbet diagram (Corbet 1986). This
diagram showed a strong correlation between their orbital and pulse
periods. An updated version of this diagram is presented in
Figure~\ref{corbet}. For a system to be detected as an X-ray source it
is necessary that the pressure of infalling material be sufficient to
overcome centrifugal inhibition. In other words, the Alfv\'{e}n radius must
be inside the magnetospheric boundary.  For BeX systems 
the correlation is understood in
terms of the wider orbits exposing the neutron stars to a lower wind
density on average and hence weaker accretion pressures. Thus it is
only those systems that have a correspondingly weaker centrifugal
inhibition (i.e. slower spin rates) that permit accretion to occur and
hence permit their possible discovery as an X-ray source.

A powerful feature of this diagram is that it allows one to predict
either the pulse, or the binary period, if one knows the other
parameter. Alternatively, if both periods are determined (from
e.g. X-ray measurements) then the class of object may be defined, even
in the absence of an optical counterpart.

\subsection{Magellanic Cloud HMXBs}

\begin{figure}
\includegraphics[width=3in]{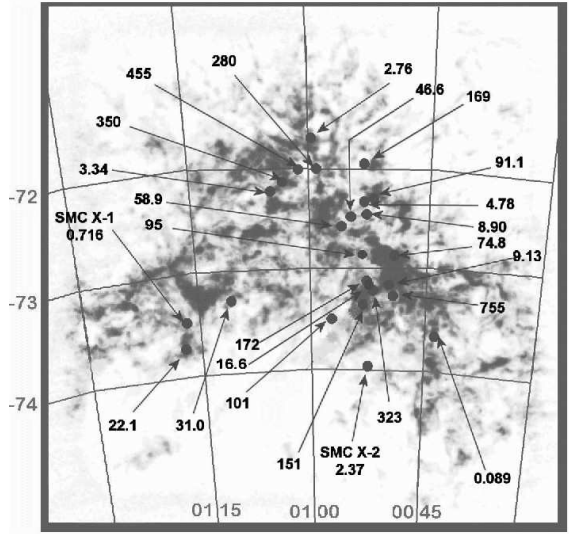}
\caption{An HI image of the SMC (Stanimirovic et
al, 1999), on which is superposed the
location of 25 known X-ray binary pulsars.  The numbers
indicate their pulse period in seconds.}\label{smc}
\end{figure}

It has come as a great surprise to discover that there are a large
number of BeX binaries in the SMC. Figure~\ref{smc} shows the
distribution of almost all the known HMXB pulsars superimposed upon an
HI image of the SMC.

It is possible to estimate the number of systems one would expect
based upon the relative masses of our Galaxy and the SMC. This ratio
is $\sim$50, so with 64 known or suspected systems in our Galaxy we
would only expect 1 or 2 in the SMC. However, Maeder et al(1999) 
have shown that the fraction of Be
to B stars is 0.39 in the SMC compared with 0.16 in our
Galaxy. So this raises the expected number of BeX systems to
$\sim$3 - but we now know of $\ge$35 (Table~\ref{smctab})!

\begin{table}
\caption{Properties of X-ray pulsars in the SMC}\label{smctab}
\begin{tabular}{llccc}
&&&&\\
\hline \\
Short	&Full or alternative			&Spectral	&V mag	&Period\\
name	&name(s)       				&Class   	&	&secs\\   

\hline
\hline

SXP0.09  &AX J0043-737        &Be?    &  &0.087 \\
SXP0.72  &SMC X-1          &B0Ib    &13.25  &0.716 \\
SXP0.92  &PSR0045-7319        &B    &16.00  &0.92 \\
SXP2.16  &XTE SMC pulsar        &    &  &2.16 \\
SXP2.37  &SMC X-2          &B1.5V    &16.00  &2.37 \\
SXP2.76  &RX J0059.2-7138      &B1III    &14.10  &2.76 \\
SXP3.34  &AX J0105-722, RX J0105.3-7210    &    &  &3.343 \\
SXP4.78  &XTE J0052-723, [MA93]537    &    &15.80  &4.782 \\
SXP5.44  &CXOU J010042.8-721132      &AXP?    &17.80  &5.44 \\
SXP7.70  &XTE SMC pulsar        &    &  &7.7 \\
SXP8.90  &RX J0051.8-7231, 1E0050.1-7247    &B1III-V  &  &8.9 \\
SXP9.13  &AX J0049-732         &    &  &9.132 \\
SXP15.3  &RX J0052.1-7319      &B1III - B0V  &14.70  &15.3 \\
SXP16.6  &RX J0051.8-7310      &    &  &16.6 \\
SXP22.1  &RX J0117.6-7330      &B0.5III  &14.20  &22.07 \\
SXP31.0  &XTE J0111.2-7317      &O8-B0V    &15.35  &31 \\
SXP46.4  &XTE SMC pulsar        &    &  &46.4 \\
SXP46.6  &1WGA 0053.8-7226, XTE J0053-724    &B1-2III-V  &14.90  &46.6 \\
SXP51.0  &XTE SMC pulsar       &    &  &51 \\
SXP58.9  &RX J0054.9-7226, XTE J0055-724   &    &  &58.95 \\
SXP74.8  &RX J0049.1-7250, AX J0049-729    &Be ?    &15.90  &74.8 \\
SXP82.4  &XTE J0052-725        &    &  &82.4 \\
SXP89.0  &XTE SMC pulsar        &    &  &89 \\
SXP91.1  &AX J0051-722, RX J0051.3-7216    &      &15.00  &91.1 \\
SXP95.2  &XTE SMC pulsar        &    &  &95.2 \\
SXP101  &AX J0057.4-7325, RX J0057.3-7325    &    &  &101.4 \\
SXP152  &CXOU J005750.3-720756, [MA93]1038  &    &  &152.1 \\
SXP164  &XTE SMC pulsar        &    &  &164.7 \\
SXP169  &XTE J0054-720, AX J0052.9-7158    &     &15.50  &169.3 \\
SXP172  &AX J0051.6-7311, RX J0051.9-7311    &    &  &172.4 \\
SXP280  &AX J0058-72.0        &     &  &280.4 \\
SXP304  &CXOU J010102.7-720658, [MA93]1240  &    &  &304.5 \\
SXP323  &AX J0051-73.3, RXJ0050.7-7316     &B0-B1V    &15.40  &323.2 \\
SXP349  &SAX J0103.2-7209, RX J0103-722    &O9-B1III-V  &14.80  &349.9 \\
SXP455  &RX J0101.3-7211      &    &  &455 \\
SXP564  &CXOU J005736.2-721934, [MA93]1020  &    &  &564.8 \\
SXP755  &AX J0049.4-7323, RX J0049.7-7323  &B1-3V    &  &755.5 \\

\hline \\

\end{tabular}
\end{table}

The answer almost certainly lies in the history of the Magellanic
Clouds. Detailed HI mapping by Stavely-Smith et al (1997)
and Putman et al (1998) has shown the existence of a
significant bridge of material between the Magellanic Clouds and
between them and our own Galaxy. Furthermore, Stavely-Smith et al have
demonstrated the existence of a large number of supernova remnants
of similar age ($\sim$5 Myr), strongly suggesting enhanced starbirth
has taken place as a result of tidal interactions between these
component systems. Consequently it seems very likely that the previous
closest approach of the SMC to the LMC $\sim$100 Myrs ago may have
triggered the birth of many new massive stars which have given rise to
the current population of HMXBs. In fact, other authors (e.g. Popov et
al 1998) claim that the presence of large numbers of
HMXBs may be the best indication of starburst activity in a system.

\begin{figure}
\includegraphics[width=3in,angle=-90]{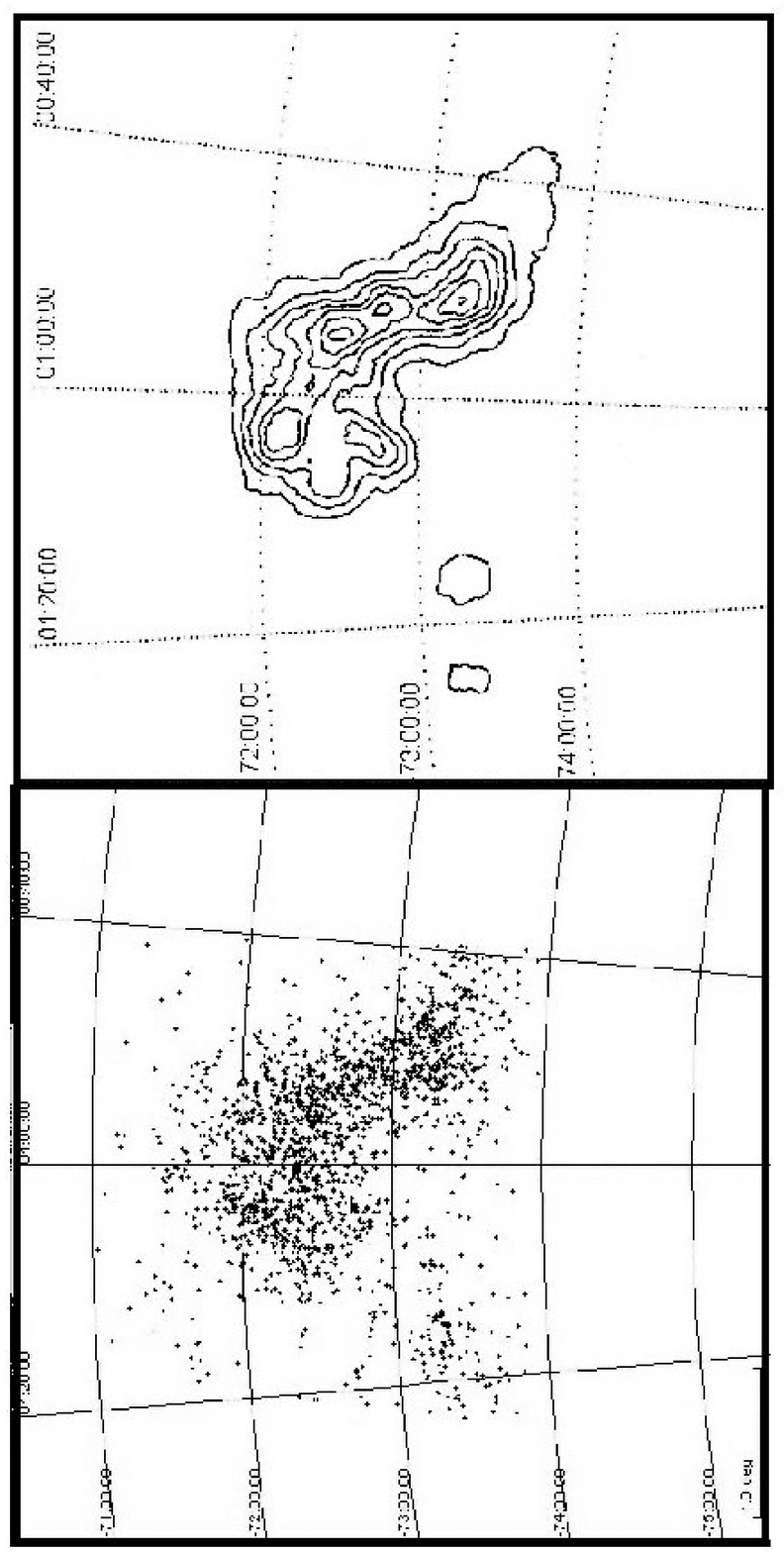}
\caption{
Left: spatial distribution of emission line stars in the
SMC Meyssonnier \& Azzopardi (1993). Right: isodensity contours for SMC stars
aged 8-12 Myr Maragoudaki et al (2001).}\label{smc2}
\end{figure}

Excellent support for this interpretation comes from both the work of
Meyssonnier \& Azzopardi (1993) in cataloguing emission
line stars, and the work of Maragoudaki et al (2001) in
identifying stars of age in the range 8-12 My. Figure~\ref{smc2} shows
the results of both of these surveys and demonstrates a strong spatial
correlation with the distribution of X-ray pulsars shown in Figure~\ref{smc}.

\begin{figure}
\includegraphics[width=2.5in,angle=-90]{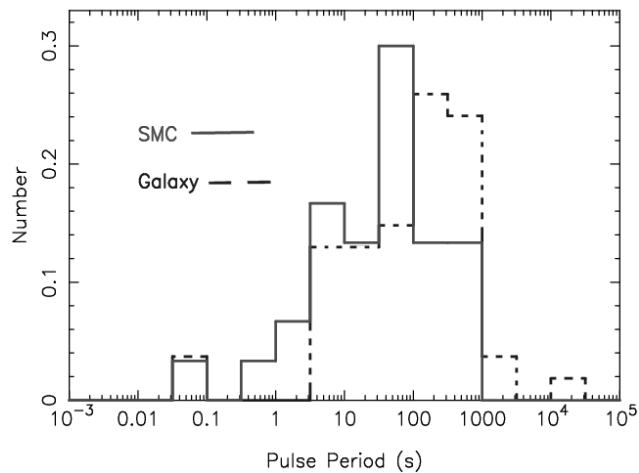}
\caption{Comparison of the X-ray pulsar population of the
Galaxy and SMC (Laycock et al, 2003).}\label{ppd2}
\end{figure}

As a result, the SMC now provides us with an
excellent sample of BeX systems in a relatively compact and
easily observable region of the sky. In addition the similarity between
the population sizes seen in the Galaxy ($\sim$50) and the
SMC ($\sim30$) permits interesting comparisons to be
made. Figure~\ref{ppd2} from Laycock et al (2003) shows
the binned population distributions per decade in period and
normalised to unity. Applying the K-S test to the two populations
gives a probability of 97\% that the two samples were drawn from
different distributions (although varying
extinction conditions may significantly affect the homogeneity of the
galactic population).

\begin{figure}
\includegraphics[width=2in]{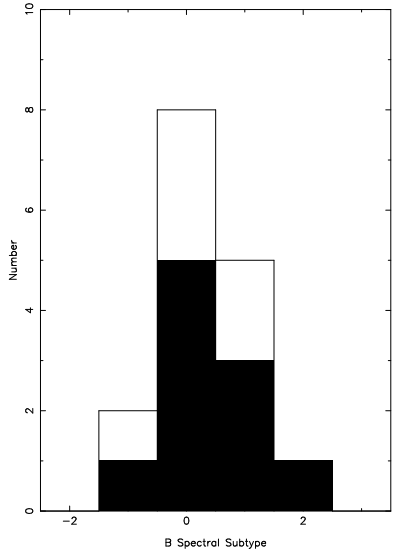}
\caption{Comparison of the X-ray pulsar optical counterparts in the
Galaxy (white) and LMC (black) Negueruela \& Coe (2002). Negative spectral
types represent O-type stars.}\label{lmc}
\end{figure}

In complete contrast, the HMXB population of the LMC seems very
representative of the sample found in the Milky Way. A comprehensive
review by Negueruela \& Coe (2002) of all identified
optical counterparts in the LMC found that the overall spectral
distribution of the population looks very similar to that of our
Galaxy (Figure~\ref{lmc}).  They found that the spectral
distribution for BeX binaries in both the Milky Way and the LMC is
sharply peaked at spectral type B0 (roughly corresponding to $M_{*}$
$\approx$16$M_{\odot}$) and not extending beyond B2
($M_{*}\approx10\,M_{\odot}$), which strongly supports the
existence of supernova kicks. So the BeX binaries in the LMC must have
preferentially formed from moderately massive binaries undergoing
semi-conservative evolution in the same manner as most galactic
systems.

\subsection{Properties of BeX circumstellar discs}

Typically the optical star exhibits H$\alpha$ line emission and
continuum free-free emission (revealed as excess IR flux) from
a disc of circumstellar gas. Therefore H$\alpha$ measurements 
provide very useful information on the disc's physical state.
A1118-616 provides an excellent example of a classic BeX system that
can be probed effectively in H$\alpha$.  Discovered by chance in 1974 
while {\it Ariel 5} was observing the nearby source Cen X-3 (Eyles et
al, 1975), the flux increased by
more than an order of magnitude during the outburst and became the
dominant hard X-ray source in the Centaurus region for 7-10 days. The
system then disappeared for 27 years before
re-emerging in the same dramatic style in 1991 (Coe et al 1994).

\begin{figure}
\includegraphics[width=1.75in,angle=-90]{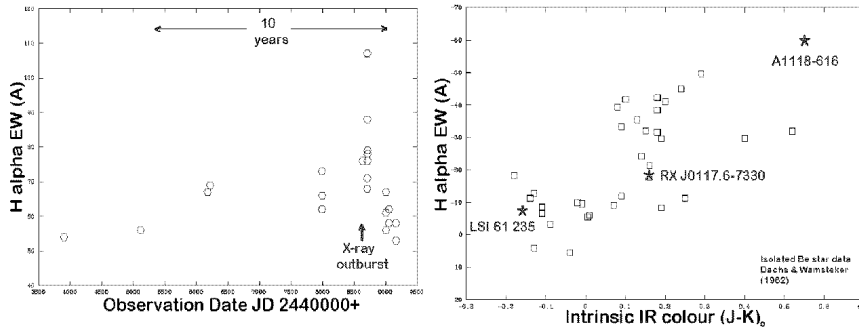}
\caption{Left - the history of the H$\alpha$ emission from A1118-616 showing
the large rise coincident with the X-ray outburst. Right - The relationship between the intrinsic IR colours and 
H$\alpha$ equivalent width. Three BeX systems are compared
to a set of isolated Be stars (Coe et al, 1994).}\label{1118bc}
\end{figure}

The explanation for these sudden massive X-ray outbursts on very long
timescales lies in the H$\alpha$ observations carried out prior to,
during, and after the 1991 outburst. Note, no similar observations
exist for the 1974 outburst since the optical counterpart was not
identified at that time. The history of the H$\alpha$ observations
over a 10 year period are shown in Figure~\ref{1118bc}, from which
it is immediately apparent why A1118-616 went into outburst
when it did - the H$\alpha$ equivalent width had reached an
exceptionally high value $\ge$100\AA.  Since the pulse period
is 404s the Corbet diagram (Corbet 1986) suggests $P_orb$ in
the range 200-300d, therefore it is extremely unlikely that the two recorded
outbursts relate to binary motion (Type I outbursts).

It is far more likely that these are Type II outbursts, and the normal
Type I outbursts are either very minor or absent. This
could simply be due to the orbit not normally taking the neutron star
through the circumstellar disc. However, during these abnormal levels
of H$\alpha$ activity, the disc probably expands to include the
neutron star's orbit, and hence accretion immediately begins. Note the
strikingly rapid decline of H$\alpha$ immediately after the
outburst suggesting that the whole period of activity was probably
just due to one major mass ejection event from the Be star.

It is interesting to compare the size of the circumstellar disc seen
in A1118-616 with that observed in other systems. Figure~\ref{1118bc}
shows a plot of H$\alpha$ EW against the intrinsic IR colour (also
thought to arise from free-free and free-bound emission in the
circumstellar disc), and an obvious and unsurprising correlation
clearly exists between these two parameters. The quiescent location of
A1118-616 is shown together with two other BeX
systems, and compared with data from a sample of isolated Be stars
(Dachs et al, 1982). It is interesting to note, that even in
quiescence, A1118-616 is at the extreme edge of the diagram, and in
fact, its peak H$\alpha$ EW value of $\sim$110\AA ~~may be one of the
very largest recorded values for any Be star.

\begin{figure}
\includegraphics[width=3in]{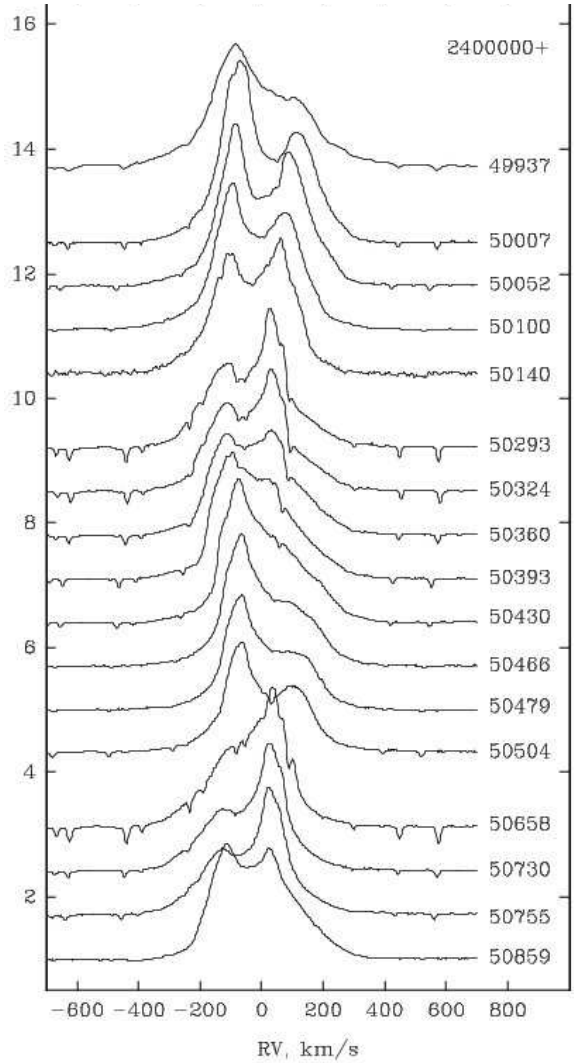}
\caption{Normalised H$\alpha$ profiles of X Per as a function of
velocity relative to the rest wavelength. 
Different dates given on the right hand side of the figure.
From Clark et al (1994).}\label{ha}
\end{figure}

With one notable exception discussed below, the H$\alpha$ profiles
follow the behavioural patterns seen in isolated Be stars. The profile shape
is dominated by the circumstellar disc structure
and its inclination to our line of sight.  A range of possible
profiles are observed and these are well documented and classified
elsewhere (eg Hanuschik et al 1988, Hanuschik
1996). A good example of a set of varying H$\alpha$
profiles from X Per is shown in
Figure~\ref{ha} (Clark et al, 2001). Not
only does this range of profiles indicate significant variations in
disc size, there is strong evidence that the blue and red components
are varying with respect to each other - known as V/R
variations. These V/R changes can only be explained by
non-uniformities in the disc structure which are rotating around the
system.

The structure of these discs has been explained by Okazaki \&
Negueruela, 2001 and others. They have investigated the
tidal torques on these circumstellar (or ``decretion'' discs and found
that a natural truncation occurs at certain resonance points (radii at
which the local Keplerian period is an integer fraction of the orbital
period). Beyond this truncation point matter cannot be
transported. Furthermore, HMXBs with very circular orbits are
truncated at a fixed size all the time, often smaller than the Roche
lobe, making accretion on to the neutron star very unlikely. As a
result these systems tend to only exhibit persistent low-level X-ray
emission from the stellar wind plus occasional Type II
outbursts. Conversely, systems with high eccentricities permit the
size of the disc to be orbital-phase dependant and during periastron
passage the disc can extend well into the orbital path of the neutron
star triggering a Type I X-ray outburst.

\begin{figure}
\includegraphics[width=1.5in]{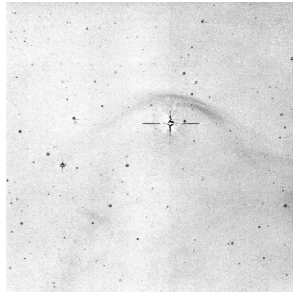}
\caption{9' x 9' H$\alpha$ image of the Vela X-1 bow shock (Kaper et 
al 1997).}\label{vela}
\end{figure}

Finally, direct H$\alpha$ imaging has shown itself to be a powerful
tool in determining the environment around HMXBs in the case of Vela
X-1. The spectacular image reported by Kaper et al 1997
(reproduced here as Figure~\ref{vela}) provides direct evidence of a
bow shock around this system. This is clear indication that Vela X-1
is travelling at supersonic velocities through the ISM.  Kaper et al
used the symmetry of the shock to derive the direction of motion and
the origin and age of the system. Their results support the Blaauw
scenario (Blaauw 1961) whereby a supernova explosion of one binary
component can result in a high space velocity of the companion (an
``OB runaway''). Surprisingly, this runaway star has a high
probability of remaining bound to its companion and thereby producing
systems like Vela X-1.

\subsection{Long-term optical monitoring}

\begin{figure}
\includegraphics[width=4cm,angle=-90]{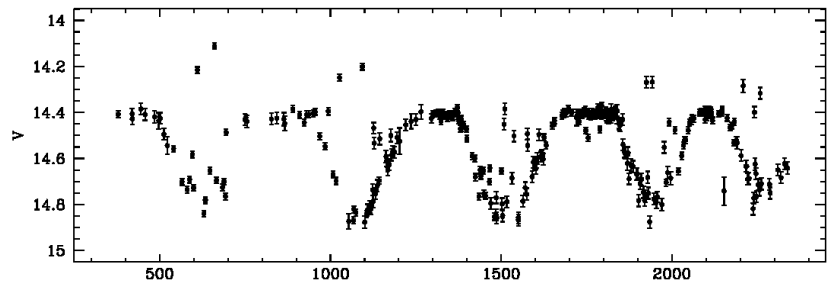}
\caption{The 5.5yr MACHO optical light curve for A0538-66 
showing a regular modulation at P=421d. 
From McGowan \& Charles (2003).}\label{0538}
\end{figure}

Of great value in the study of BeX systems has been the availability
of long term observational archives such as OGLE (Udalski et al, 1997)
and MACHO (Alcock et al, 1996). 
Such databases have permitted the long term optical
variability to be studied, sometimes in conjunction with similar
timescale X-ray databases. Two excellent examples of the science
achievable from these optical archives are the results on A0538-66
(Alcock et al 2001) and AX J0051-733 (Coe et al, 2002). 
In the case of A0538-66
the MACHO data are shown in Figure~\ref{0538} and clearly indicate a
very regular modulation of $\sim$420d, but also evidence for activity
on the much shorter 16.6d binary period (already well-known from X-ray
data, Skinner 1980, 1981). Remarkably, the shorter period outbursts
{\it only} occur during the {\it minima} in the 420d cycle. McGowan \&
Charles 2003 interpret this as being due to evolution in
the circumstellar disc.

\begin{figure}
\includegraphics[width=1.5in, angle=-90]{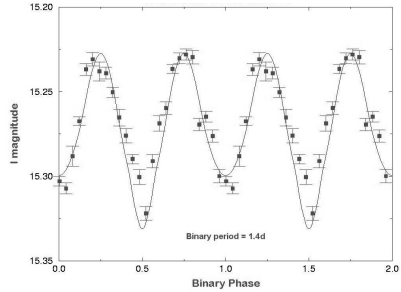}
\caption{Comparison of a representative ellipsoidal model ($i=60^{\circ}$,
$Q=0.20$, $f=0.99$, and $r_d=0.33$) and the folded I-band data for RX
J0050.7-7316.  The amplitude and relative depths of the minima are
roughly matched by the model.  However, there are relatively large
deviations, especially between phases 0.4 and 0.6 
(Coe \& Orosz 2000).}\label{0051}
\end{figure}

The second example of a source studied using these kinds of data is AX
J0051-733.  A combination of MACHO and OGLE data (Coe et al, 2002)
revealed an extremely short modulation period of either 0.7d or double
that value - see Figure~\ref{0051}. The shape of this optical
modulation is driven both by tidal distortion of the envelope of the
mass donor (see e.g. Avni \& Bahcall 1975 and Avni 1978), and by X-ray
heating effects (Orosz \& Bailyn 1997). Orbital parameters such as
inclination and mass ratio, may be determined from these light curves.

\subsection{Spectral classification of the mass donors}

The evolutionary history of Be stars in HMXBs is somewhat
different to that of their isolated colleagues.  The widely accepted
evolutionary path is based upon conservative mass transfer 
developed by van den Heuvel (1983) and
Verbunt \& van den Heuvel (1995). The important
consequence of this scenario is that wide binary orbits (200 -- 600d)
are produced before the final supernova explosion. Hence, any small
asymmetries in the subsequent SN explosion will then produce the
frequently observed wide eccentric orbits.

\begin{figure}
\includegraphics[width=4cm]{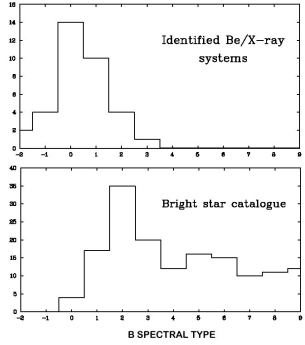}
\caption{Comparison of the spectral classes of Be stars in BeX
binaries (upper) with those in isolated systems (lower). 
Adapted from Negueruela (1998).
}\label{comp}
\end{figure}

Of particular interest is the narrow range of spectral class revealed
through blue spectroscopy of the identified counterparts (see
Negueruela 1998). In contrast to the sample of isolated
Be stars found in the Bright Star Catalogue, there are no known
BeX objects beyond spectral class B3 - see
Figure~\ref{comp}. Most commonly these have counterparts in the
B0-B2 group.

The explanation offered by Van Bever and Vanbeveren
(1997) for this phenomenom is that the wide orbits
produced by the evolutionary models are very vulnerable to disruption
during the SN explosion. This will be particularily true for the less
massive objects that would make up the later spectral classes. Hence
the observed distribution confirms the evolutionary models.

\begin{figure}
\includegraphics[width=7.5cm]{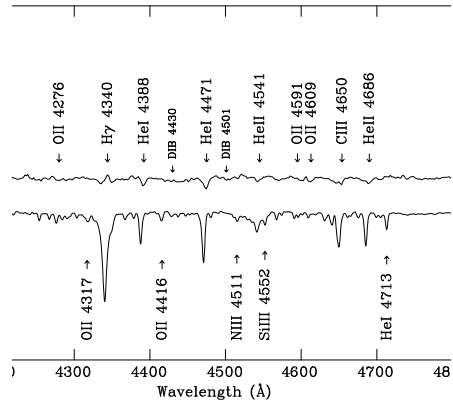}
\caption{
Spectrum of the optical counterpart of RX J0052-7319 (type B0Ve) in the
classification region compared to the B0V standard $\nu $
Ori (Covino et al, 2001)}.
\label{spect}
\end{figure}

An excellent example of using optical spectroscopy to investigate an
HMXB is the work of Covino et al (2001) on a pair of
SMC objects. The blue spectrum of RX J0052-7319 (Figure~\ref{spect}) 
is compared to a standard star of the same spectral
type, where it is clear that some of the weaker He I lines are not
visible, presumably filled in by emission. All the spectral features
are very shallow and broad, which is typical of a BeX
counterpart and the presence of weak He II absorption places it 
close to B0.  Furthermore, if the object is on the main
sequence, the presence of weak HeII $\lambda4200$\AA ~and the
condition HeII $\lambda4541$ $\>>$ SiIII $\lambda4552$\AA ~give a
spectral type of B0Ve.

\subsection{IR spectroscopy of HMXBs}

The observations summarised in earlier sections reveal that many of
the counterparts to HMXBs exhibit mass outflow to the extent of
creating a circumstellar disc of material around the mass
donor. Free-free and bound-free IR emission from this disc show
themselves as a significant excess over the normal stellar spectrum at
all wavelengths greater than the V band. This IR signature, often
quantified as a J-K colour excess, is important for the following
reasons:

\begin{itemize}

\item in confirming the identity of a Be star in the absence of
optical spectral information;

\item in providing an estimate of the size of the circumstellar disc
(this is often directly related to the magnitude of the X-ray emission);

\item through IR spectroscopy providing a channel for spectral
identification in the case of objects suffering heavy optical extinction.

\end{itemize}

\begin{figure}
\includegraphics[width=4cm,angle=-90]{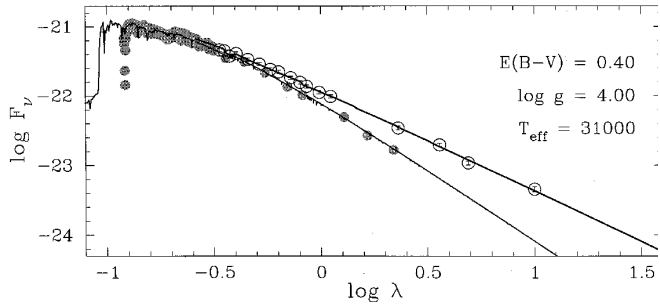}
\caption{
Comparison of the optical/IR emission from X Per in two different
states: with (upper curve) and without 
a substantial circumstellar disc Telting et al (1998).
}\label{xper}
\end{figure}

One system that has been the subject of extensive IR observations over
the years is X Per. Detailed study by
Telting et al (1998) found that the density of the disc
varies along with the brightness of X Per, and that in optical high
states the disc in X Per is among the densest of all Be stars:
$\rho_0$=(1.5$\pm$0.3)x10$^{-10}$ g cm$^{-3}$. The disc density at the
photosphere varies by a factor of at least 20 from optical high to low
states (see Figure~\ref{xper} for large differences taking place in
the IR).

\begin{figure}
\includegraphics[width=8cm]{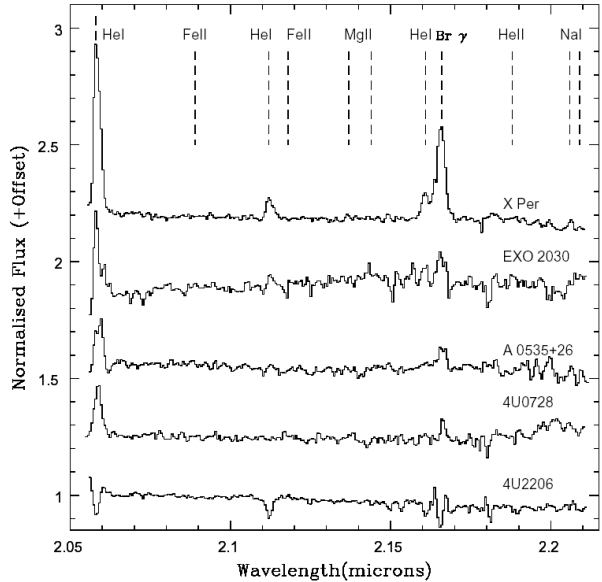}
\caption{
K band spectra of BeX binaries (Clark et al, 1999). The
positions of prominent H, He and metallic transitions are
marked.  }\label{irs}
\end{figure}

With the recent advent of IR spectroscopy on large telescopes, the
possibilities of obtaining spectra despite high levels of extinction
have become a reality. This is a very powerful tool for exploring
previously inaccesible objects, but also directly gleaning information
on the circumstellar environment. A good example of the strength of
this new tool may be seen in Clark et al 1999 which shows
the annotated IR spectra of 5 HMXBs (Figure~\ref{irs}). Just as the
optical spectrum is dominated by Balmer emission, we see the same
effect from the Brackett series emission in the IR. In addition,
emission lines from He and metallic transitions are
prominent. This approach, therefore, offers great potential for
identifying and classifying the more optically obscured systems, as
well as revealing the state of the circumstellar disc.

\subsection{UV spectroscopy of HMXBs}

\begin{figure}
\includegraphics[width=8cm]{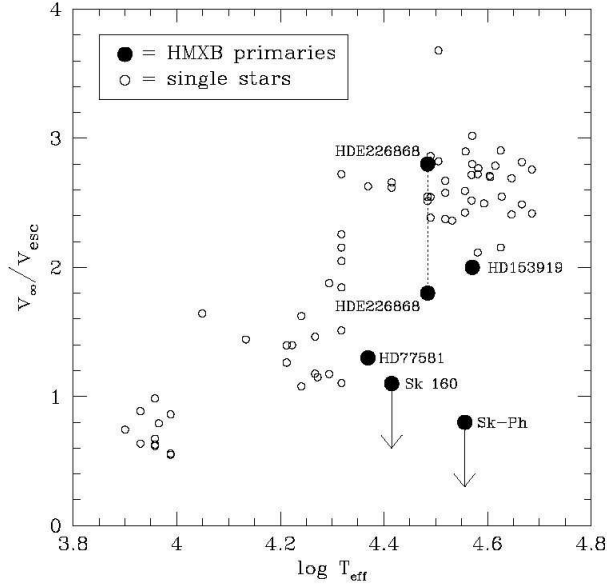}
\caption{
The ratio of terminal over escape velocity for the mass donor stars in
5 HMXBs compared to isolated stars. The HMXBs have low terminal
velocities for their effective temperatures van Loon et al(2001).
}\label{iue}
\end{figure}

UV spectroscopy of the mass donors in HMXBs has proved a powerful
tool for understanding these systems. Since these are hot
young stars they tend to be extremely UV bright. Important information about the stellar outflow may be
determined from P Cygni profiles, as well as direct estimates of
interstellar absorption from the 2200\AA~ feature which 
is valuable in de-coupling local and ISM extinction.

van Loon et al(2001)  carried out
a comprehensive study of 5 HMXBs using extensive IUE UV data. 
Using radiation transfer codes they derived the terminal wind 
velocities and sizes of the Stromgren
zones created by the X-ray source from modelling the UV resonance
lines. They found that these sizes were in good
agreement with that expected from standard Bond-Hoyle accretion
and scaled with $L_X$. However, the determined terminal
velocities were lower than expected from single stars with the same
$T_eff$ - see Figure~\ref{iue}. These lower velocities
could be due to the presence of the X-ray Str\"{o}mgren sphere inhibiting the
wind acceleration.

In addition, blue and UV spectra are essential tools for determing the
precise spectral class of the mass donor, and hence information about
the compact object. Clark et al 2002 used high resolution
IUE UV spectra of 4U1700-37 to carry out detailed
non-LTE modelling of the O type primary. Using the modelling code of
Hillier \& Miller 1998 which utilises radiative transfer
processes in a spherical extended atmosphere, Clark et al were able to
precisely define the mass donor star to be of spectral type
O6.5Iaf. Combining this with binary information from
eclipse monitoring, led to an accurate determination of an unusually
high neutron star mass of 2.44$\pm$0.27\msun. This is one of the
highest accurate mass determinations for a neutron star, but is not
alone in pushing upwards the mass of neutron stars in HMXBs. Barziv et
al 2001 have 1.87$\pm$0.20\msun~for
the compact object mass in Vela X-1, whereas Orosz \& Kuulkers
1999 find a value of 1.78$\pm$0.23\msun~for Cyg X-2.
However, it is still conceivable that there are systematic effects at
work here given how the neutron star ``mass'' correlates with spectral
type.

\section{\sc {Low-mass X-ray binaries \label{sect:LMXBs}}}

\bigskip

The review by van Paradijs \& McClintock (1995) remains an excellent
introduction to the optical characteristics of both HMXBs and LMXBs,
and Liu et al (2000, 2001) provide comprehensive catalogues of their
observed properties.  However, this is a field that has extended both
in wavelength (with high quality UV spectroscopy available from
HST/STIS, and cooled grating spectrographs allowing IR spectroscopy)
and sensitivity (the advent of large telescopes such as Keck, VLT,
Subaru and VLT) during the last decade.  Consequently this section
will focus on these later developments, and in particular on the
sub-class of {\it X-ray transients} which have allowed us to gain a
far more detailed knowledge of the properties of the mass donor in
LMXBs than was possible a decade ago.  Additionally, gains in CCD
technology now permit high time resolution photometry and spectroscopy
and a consequential improvement in our understanding of luminous
accretion discs.

\subsection{Fundamental Properties}

The presence of a ``steady'' X-ray source implies that the companion
fills its Roche lobe, and so (Paczynski 1971)

\begin{equation}
R_2/a = 0.46(1+q)^{-1/3} 
\end{equation}

where the mass ratio $q = M_X/M_2$.  Combined with Kepler's 3rd Law,
then

\begin{equation}
{\rho} = 110/P^2_{hr}   
\end{equation}

where $\rho$ (in g~cm$^{-3}$) is the mean density of the secondary.
\footnote{N.B. the Paczynski
relation is only valid for $q>1$, and there is a more accurate
algorithm (Eggleton 1983) which is valid for all $q$.}

These stars are presumed to be on or close to the lower main sequence,
then $M_2=R_2$ and hence $M_2=0.11P_{hr}$ (a detailed observational
analysis of the secondary stars in CVs and LMXBs can be found in Smith
\& Dhillon 1998).  Hence short period X-ray binaries must
be LMXBs, their companion stars will be faint, and the optical light
will be dominated by reprocessed X-radiation from the disc (or heated
face of the companion star, although this is likely substantially
shielded by the disc itself; see van Paradijs \& McClintock
1994).  This obliteration of the companion in ``steady''
LMXBs removes one of the key pieces of dynamical information about the
binary.  The optical spectra of LMXBs consist of hot, blue continua
($U-B$ typically $-1$) on which are superposed broad H, He emission
lines, with velocity widths typical of the inner disc region
($\sim$500--1000 km~s$^{-1}$), making accurate velocity information
about the motion of the compact object difficult to obtain.
Consequently accurate dynamical information about both components in
LMXBs has been very difficult to acquire, yet is essential if accurate
masses are to be determined (see next section).  Hence, evidence on
the nature of the compact object in most bright LMXBs is indirect,
e.g. X-ray bursting behaviour for neutron stars (as few are X-ray
pulsars) or fast variability as first seen in Cyg X-1 (and used as a
possible black hole diagnostic, see chapter 4).

\subsection{Dynamical mass measurements in X-ray novae}

Progress in identifying the nature of compact objects in LMXBs
requires dynamical mass measurements of the type hitherto employed on
X-ray pulsars in HMXBs (see e.g. White et al 1995).
Unfortunately, all that can be measured (in the case of Cyg X-1, and
the other two HMXBs suspected of harbouring black holes, LMC X-1 and
LMC X-3) is the mass function

\begin{equation}
f(M) = {{PK^3}\over {2{\pi}G}} = {{M^3_X{\sin}^3i}\over{(M_X+M_2)^2}}
\end{equation}

where $K$ is the radial velocity amplitude.  With $M_2{\geq}M_X$, then
$M_X$ is poorly constrained due to the wide range of uncertainty
($\sim$12--20M$_\odot$) in $M_2$ as a result of their unusual
evolutionary history.  

\begin{figure}[t]
\begin{center}
\begin{picture}(100,250)(50,30)
\put(0,0){\includegraphics{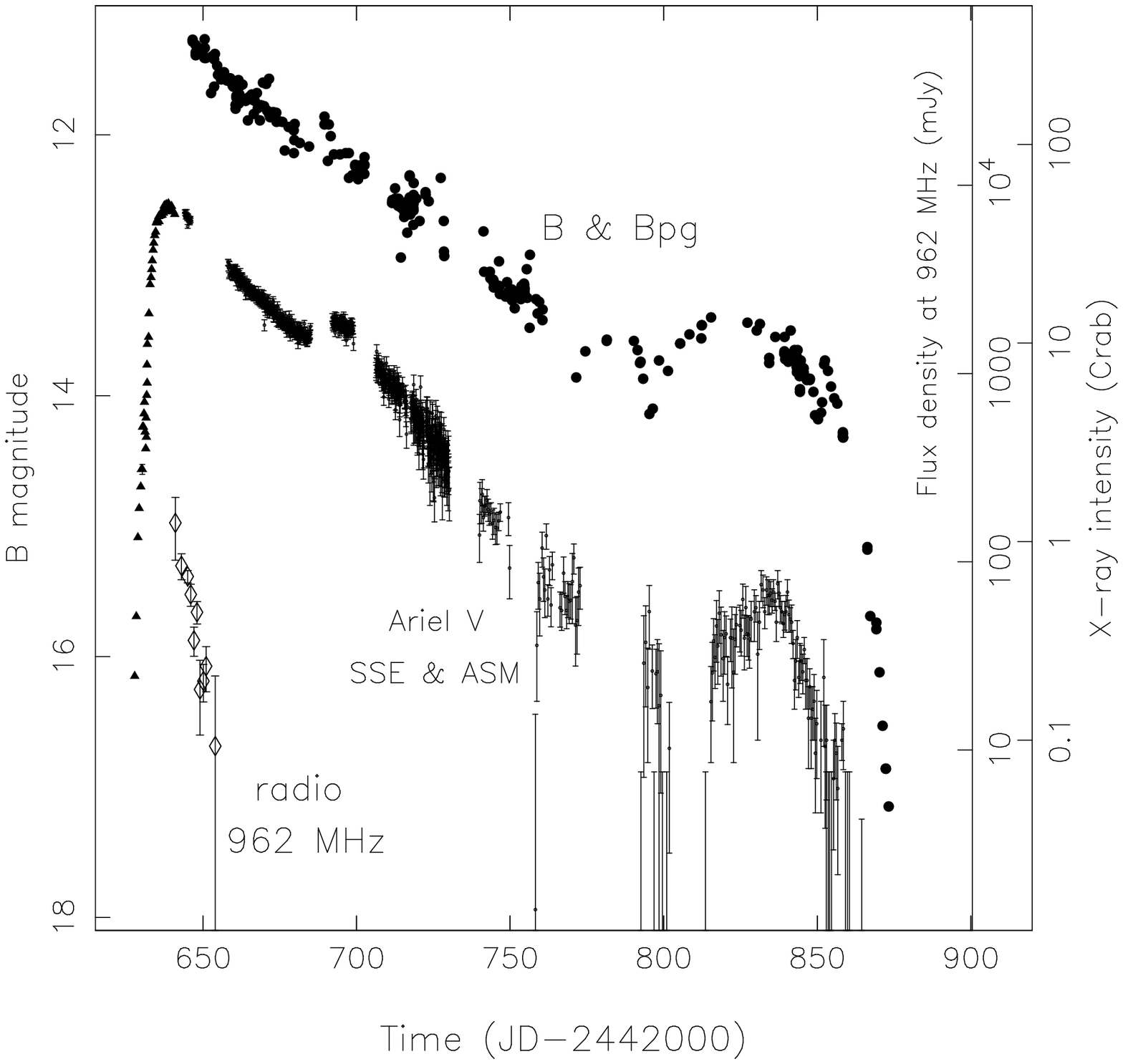}}
\noindent
\end{picture}
\caption{Optical, X-ray and radio outburst light curves of the
prototype XRN A0620-00 (=Nova Mon 1975) showing the typical fast rise,
exponential decay on a timescale of months, adapted from Kuulkers
(1998). \label{a0620}}

\end{center}
\end{figure}

LMXB transients (usually referred to as {\it soft X-ray transients
(SXTs)} or {\it X-ray novae (XRN)} provide ideal opportunities to
study the properties of the secondary stars in great detail.  Detected
in outburst by X-ray all-sky monitors, they usually fade within
timescales of months, after which the companion star becomes visible.
The prototype of this sub-class is A0620-00, still the brightest of
all XRN when it erupted in 1975, the data from which were
comprehensively re-examined by Kuulkers (1998) producing the revised
light curve shown in figure~{\ref{a0620}}.  Currently $\sim$25\% of
XRN are confirmed neutron star systems (they display type I X-ray
bursts), the remainder are all black-hole candidates, the highest
fraction of any class of X-ray source (see e.g. Chen et al 1997).
Their X-ray properties are discussed in chapter 4, and their
optical/IR properties are summarised in table~{\ref{tab:SXTs}}.

\begin{sidewaystable}
\begin{center}
\tiny
\label{tab:SXTs}
\begin{tabular}{lcccccccccccccccc} 
\hline
{\em X-ray Source} &  {\em Optical/IR} & {\em b$^{II}$} & {\em Outbursts} & {\em P} & {\em Sp.} &  
{\em E$_{B-V}$} & {\em V} & {\em K$^*$} & 
{\em v$\sin$i} & {\em K$_2$} & $\gamma$ &{\em f(M)} & 
{\em q} & {\em i} & {\em M$_X$} & {\em M$_2$} \\
 &  {\em counterpart} & & & {\em (hrs)} & {\em Type} &  &\multicolumn{2}{c}{\em (quiesc)} & 
\multicolumn{3}{c}{\em (km~s$^{-1}$)} & ($M_\odot$) &  (={\em M$_X$/M$_2$}) & ($^o$)  & ($M_\odot$) & 
($M_\odot$)\\  
\hline
\multicolumn{2}{c}{\bf BLACK HOLES} & \\
& & & & & \\
{\it {\bf LMXBs}} & \\
GRS1915+105$^{\bf 1}$ & V1487 Aql & -0.2 & 1991+``on'' & 816 & K III & $\sim$10 & - & 13 & - & 140 & -3 & 9.5$\pm$3 & 12 & 70$\pm$2 & 14$\pm$4 & 1.2$\pm$0.2 \\
J1859+226$^{\bf 2}$ & V406 Vul & +8.6 & 1999  & 9.2 & - & 0.58 & 23.3 & - & - & 570 & - & 7.4$\pm$1.1 & - & - & $>$5 & - \\
J1550-564$^{\bf 3}$ & V381 Nor & -1.8 & 1998, 00, 03 & 37.0 & G8-K4IV  & 1.6 & 22 & - & 90$\pm$10 & 349 & -68 & 6.9$\pm$0.7 & - & 72$\pm$5 & 9.6$\pm$1.2 & - \\
J1118+480$^{\bf 4}$ & KV UMa & +62.3 & 2000 & 4.1 & K7-M0 & $<$0.02 & 19.6 & & 114$\pm$4 & 701 & -15 & 6.1$\pm$0.3 & $\sim$20 & 81$\pm$2 & 6.8$\pm$0.4 & 0.25$\pm$0.15  \\
GS2023+338$^{\bf 5}$  &  V404 Cyg & -2.2 & 1938,56,89 & 155.3 & K0IV &  1 & 18.4 & 12.5 & 39$\pm$1 & 208.5 & 0 & 6.08$\pm$0.06 &  17$\pm$1 & 55$\pm$4 & 12$\pm$2 &
0.6 \\
GX339-4$^{\bf 6}$ & V821 Ara & -4.3 & frequent & 42.1 & - & 1.1 & $\sim$21 & -  & - & 317$^{\bf \dagger}$ & $\sim$+30 & 5.8$\pm$0.5 & - & - & $>$2.0 & - \\
GS2000+25$^{\bf 7}$  & QZ Vul & -3.1 & 1988 & 8.3 & K5V & 1.5  & 21.5 & 17  & 86$\pm$8 & 520 & +19 & 4.97$\pm$0.10 &  24$\pm$10 & 56$\pm$15  &
10$\pm$4 & 0.5 \\
H1705-25$^{\bf 8}$  & V2107 Oph  & +9.1 & 1977 & 12.5 & K & 0.5 & 21.5 & - & $\leq$79  & 441 &-54 &  4.86$\pm$0.13 &  $>$19 & 60$\pm$10 & 6$\pm$2
& 0.3  \\
GRS1009-45$^{\bf 9}$ & MM Vel & +9.3 & 1993 & 6.8 & K7-M0  & 0.2 & $\sim$22  & & & 475 & +41 & 3.17$\pm$0.12 & 7$\pm$1 & 67$\pm$3 & 5.2$\pm$0.6 & 0.7 \\
N Mus 91$^{\bf {10}}$  & GU Mus  & -7.1 & 1991 & 10.4 & K0-4V &  0.29 & 20.5 & 16.9 & 106$\pm$13 & 421 & +20 & 3.34$\pm$0.15 &   6.8$\pm$2 &
54$^{+20}_{-15}$ & 6$^{+5}_{-2}$ & 0.8  \\
A0620-00$^{\bf {11}}$  & V616 Mon & -6.5 & 1917,75 & 7.8 & K5V &   0.35 & 18.3 &  14.5 &83$\pm$5 & 433 & +4 & 2.91$\pm$0.08 &  15$\pm$1 & 37$\pm$5 & 10$\pm$5 & 0.6  \\
J0422+32$^{\bf {12}}$   & V518 Per  & -11.9 & 1992 & 5.1 & M1V & 0.25   & 22.3 & 17.4 & 90$\pm$25 &  372 & +11 & 1.19$\pm$0.02 &  9.0$^{+2.2}_{-2.7}$ & 45$\pm$2 &
4$\pm$1 & 0.3 \\
 & & & & &   \\
{\it {\bf IMXBs}} & \\
J1819.3-2525$^{\bf {13}}$ & V4641 Sgr & -4.8 & 1999, 02 & 67.6 & B9III & 0.32 & 13.7 & & 99$\pm$2 & 211 & +107 & 3.13$\pm$0.13 & 2.31$\pm$0.08 & 75$\pm$2 & 7.1$\pm$0.3 & 3.1$\pm$0.3  \\
J1655-40$^{\bf {14}}$  & V1033 Sco & +2.5 & 1994-96 & 62.9 & F6IV & 1.2 & 17.2 & - & 88$\pm$5 & 228 & -148 & 2.73$\pm$0.15 &  2.39$\pm$0.15 & 70$\pm$2 & 6.6$\pm$0.5 & 2.8$\pm$0.3  \\
4U1543-47$^{\bf {15}}$ & IL Lup & +5.4 & 1971,83,92,02 & 27.0 & A2V & 0.5 & 16.6 & - & - & 124 & -87 & 0.25$\pm$0.01 &  3.6$\pm$0.4 & 21$\pm$2 & 9.4$\pm$1 & 2.5  \\
 & & & & &  \\
{\it {\bf HMXBs}} & \\
LMC X-3$^{\bf {16}}$ & *1 & -32.1 & ``steady'' & 40.8 & B3Ve & 0.06 & 17.2 & - & 130 & 235 & +310 & 2.3$\pm$0.3 & 2.2--4.4 & 67$\pm$3 & 4--7 & 1--3\\
Cyg X-1$^{\bf {17}}$ & HDE 226868 & +3.1 & ``steady'' & 134.4 & O9.7Iab & 1.06 & 8.9 & - & 155 & 74.9 & -1 & 0.244$\pm$0.005 & 0.4--0.8 & 27--67 & $>$4.8 & $>$11.7 \\
LMC X-1$^{\bf {18}}$ & *32 & -31.5 & `steady'' & 101.3 & O8III & 0.37 & 14.5 & - & $\sim$150 & 68 & +221 & 0.14$\pm$0.04 & $\sim$0.5 & 2.5--6 & 8--20 \\
 & & & & &  \\
\multicolumn{2}{c}{\bf NEUTRON STARS} & \\
 & & & & &  \\
J2123-058$^{\bf {19}}$ & LZ Aqr  & -36.2 & 1998 & 6.0 & K7V & 0.12 & 21.8 & - & 121 & 299 & -95 & 0.68$\pm$0.05 & 2.7$\pm$1.0 & 73$\pm$4 & 1.5$\pm$0.3 & 0.5$\pm$0.3  \\ 
Cen X-4$^{\bf {20}}$  & V822 Cen  & +23.9 & 1969,79 & 15.1 & K3-5V &   0.1  & 18.4 & 14.8 & 43$\pm$6 & 150 & +184 & 0.22$\pm$0.01 &  5.9$\pm$1.6 & 43$\pm$11 &  1.5$\pm$1.0 & 0.3$\pm$0.2  \\  
\hline

\end{tabular}

\end{center}

{\footnotesize References: $^{\bf *}$ mostly from Wachter 1998; 
$^{\bf 1}$Greiner et al 2001; 
$^{\bf 2}$Filippenko \& Chornock 2001;
Hynes et al, 2002b; Zurita et al 2002b;
$^{\bf 3}$Orosz et al 2002a; 
$^{\bf 4}$Wagner et al 2001; Hynes et al 2000; Orosz 2001; Zurita et al 2002a;
$^{\bf 5}$;Casares et al 1992; Casares \& Charles 1994; Shahbaz et al 1994b; 
$^{\bf 6}$Hynes et al 2003b;$^{\bf \dagger}$Bowen fluorescence velocity, see section~{\ref{sect:LumLMXBs}};
$^{\bf 7}$Filippenko et al 1995a; Beekman et al 1996; Harlaftis et al 1996; 
$^{\bf 8}$Filippenko et al 1997; Remillard et al 1996; Martin et al 1995; Harlaftis et al 1997; 
$^{\bf 9}$della Valle et al 1997; Filippenko et al 1999; Gelino 2003;
$^{\bf {10}}$Orosz et al 1996; Casares et al 1997; Shahbaz et al 1997; Gelino et al 2001a; 
$^{\bf {11}}$McClintock \& Remillard 1986; Orosz et al 1994; Marsh et al 1994; Shahbaz et al 1994a; Gelino et al 2001b; 
$^{\bf {12}}$Filippenko et al 1995b; Beekman et al 1997; Harlaftis et al 1999; Webb et al 2000; Gelino \& Harrison 2003;
$^{\bf {13}}$Orosz et al 2001; Orosz 2003;  
$^{\bf {14}}$Orosz \& Bailyn 1997; van der Hooft et al 1997, 1998; Hynes et al 1998; Shahbaz et al
1999b, 2000; Greene et al 2001; Beer \& Podsiadlowski 2002; Shahbaz 2003;
$^{\bf {15}}$Orosz et al 2002b;
$^{\bf {16}}$Kuiper et al 1988;
$^{\bf {17}}$Herrero et al 1995; Brocksopp et al 1999;
$^{\bf {18}}$Hutchings et al 1983, 1987;
$^{\bf {19}}$Tomsick et al 2001, 2002; Hynes et al 2001a; Casares et al 2002; Shahbaz et al 2003; 
$^{\bf {20}}$McClintock \& Remillard, 1990; Shahbaz et al 1993; Torres et al 2002
}
\end{sidewaystable}

\normalsize

\subsection{Quiescence studies of X-ray novae \label{sect:qXRN}}

In quiescence, XRN become a valuable resource for research into the
nature of LMXBs.  Typically their optical brightness has declined by a
factor of $\geq$100, with all known XRN having quiescent
V$\sim$16--23, which is now dominated by the companion. With large
telescope optical/IR spectroscopy it is possible to determine its
spectral type, period and radial velocity curve, the latter two giving
$f(M)$ (equation 5.3), see table~{\ref{tab:SXTs}}.  These are
separated into 3 sections: BHCs with cool secondaries, those with
earlier spectral types and the neutron star XRN.
Table~{\ref{tab:SXTs}} shows the significance of such work, since all
are LMXBs with $M_X>M_2$.  N.B. $f(M)$ 
represents the {\it absolute minimum} values for
$M_X$ since we must have $i<90^o$ and $M_2>0$.

Some dynamical information can be derived even during outburst, given
spectroscopic data of sufficient resolution.  Casares et al (1995)
observed GRO~J0422+32 during an X-ray mini-outburst and found intense
Balmer and HeII$\lambda$4686 emission modulated on what was
subsequently found to be $P_{orb}$.  Furthermore, a sharp
component within the (highly complex) HeII emission profile displayed
an S-wave that was likely associated with the accretion hotspot.

However, to determine $M_X$, additional constraints are needed in
order to infer values for $M_2$ and $i$.

\subsubsection{Rotational broadening of companion spectrum}

In short period interacting binaries the secondary is constrained to
corotate with the primary and must also be filling its Roche lobe (in
order for mass transfer to occur).  Hence 
$R_2$ is given by equation (5.1), and then (Wade and Horne,
1988)

\begin{equation}
v_{rot}\sin i = {{2{\pi}R_2}\over{P}}{\sin}i = K_2\times
0.46{(1+q)^{2/3}\over{q}} 
\end{equation}

from which $q$ can be derived if $v_{rot}$ is measured.  Typically
$v_{rot}\sim$ 40--100\kms, which requires high resolution and high
signal-to-noise spectra of the secondary.

\begin{figure}[t]
\begin{center}
\begin{picture}(100,250)(50,30)
\put(0,0){\includegraphics{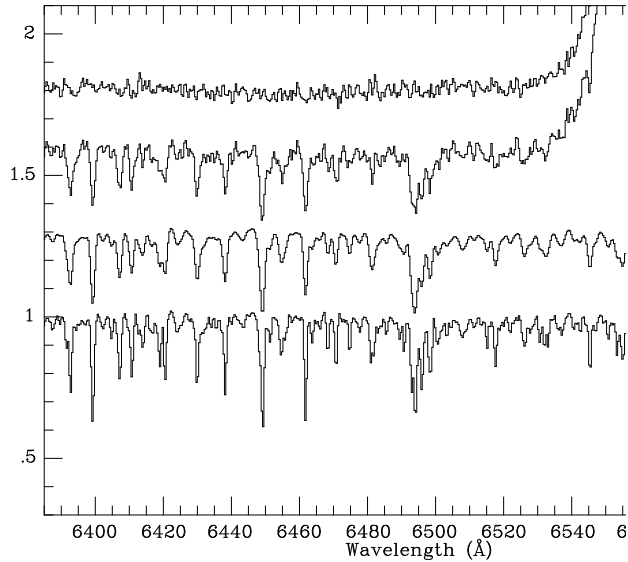}}
\noindent
\end{picture}
\caption {Determining the rotational broadening in
V404 Cyg. From bottom to 
top: the K0IV template (HR8857); the same spectrum broadened by 39 
km s$^{-1}$; doppler corrected sum of V404 Cyg (dominated by disc 
H$\alpha$ emission); residual after subtraction of 
the broadened template (from Casares and Charles, 1994).
\label{rot}}
\end{center}
\end{figure}

The technique is demonstrated in figure~{\ref{rot}} for V404 Cyg
(Casares \& Charles 1994) and based on Marsh et al (1994).  The
doppler-corrected and summed spectrum of V404 Cyg is dominated by
broad H$\alpha$ emission from the disc, but the cool companion
absorption features are visible and clearly broader than those of the
template K0IV spectrum.  The template is then broadened by different
velocities (together with the effects of limb darkening), subtracted
from that of V404 Cyg and the residuals $\chi^2$ tested, giving
$v_{rot}\sin i = 39\pm1$ \kms and hence $q$ = 16.7$\pm$1.4.  Note that
small radial velocity amplitudes have also been seen in the H$\alpha$
emission (e.g. Orosz et al 1994; Soria et al 1998), but their
interpretation is complex as there is a small phase offset
($\simeq$0.1) relative to the companion (somewhat larger in
GRS1009-45, see Filippenko et al 1999), and so until this is fully
understood, it cannot be used to determine $q$.

\begin{figure}[bt]
\begin{center}
\begin{picture}(100,250)(50,30)
\put(0,0){\includegraphics{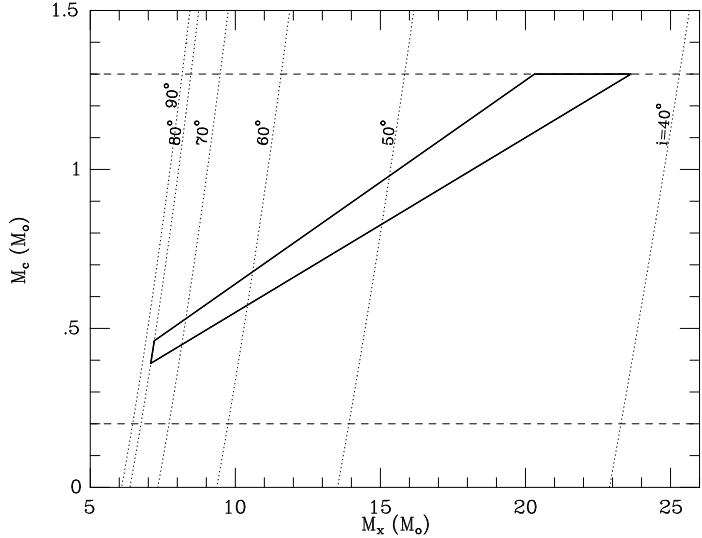}}
\noindent
\end{picture}
\caption {Constraints on $M_X$ and $M_2$ for a range
of values of $i$ in V404 Cyg based on the radial 
velocity curve ($f(M)$=)6.1\msun and determination of $q$ (=16.7, from rotational 
broadening).  It is the limited constraint on $i$ (absence of eclipses) 
that leads to a wide range of $M_X$ (Casares and Charles, 1994).
\label{constraints}}
\end{center}
\end{figure}

The combination of $q$ and $f(M)$ then yields the mass constraints
(fig~{\ref{constraints}}), and the only remaining unknown is $i$.  To
date, none of the SXTs is eclipsing (although GRO~J1655-40 and
XTE~J2123-058 show evidence for grazing eclipses), and so it is the
uncertainty in $i$ that dominates the final mass measurement.
Nevertheless there are methods by which $i$ can be estimated.

\subsubsection{Ellipsoidal modulation}

\begin{figure}[t]
\begin{center}
\begin{picture}(100,250)(50,30)
\put(0,0){\includegraphics{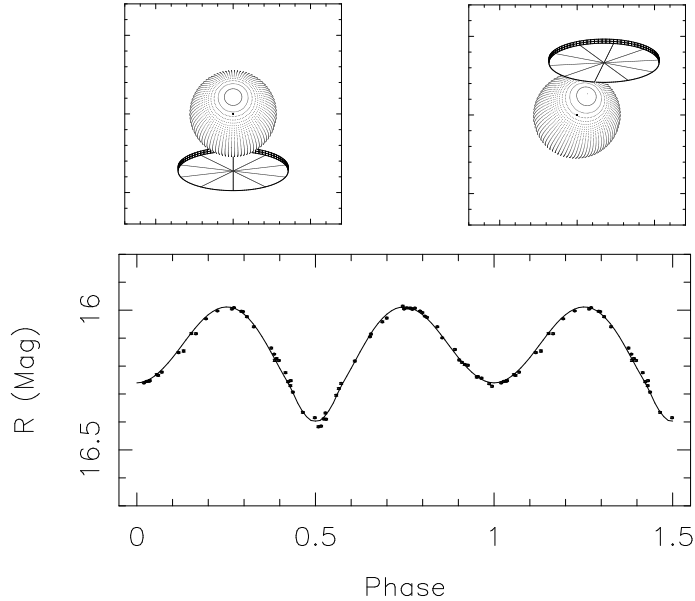}}
\noindent
\end{picture}
\caption {R-band light curve of GRO J1655-40 in quiescence 
(van der Hooft et al 1997) together with schematics of the system
orientation (Orosz \& Bailyn 1997) which show the grazing eclipses
required by the fits to the light curve. \label{J1655lc}}
\end{center}
\end{figure}

The secondary star in interacting binaries has a peculiarly distorted
shape which gives rise to the so-called {\it ellipsoidal modulation}
due to its varying projected area as viewed around the orbit.  This
leads to the classical double-humped light curve
(fig~{\ref{J1655lc}}), which is well defined by theory (i.e. the form
of the Roche lobe) and the observed light curve depends principally on
$q$ and $i$.  For large $q$ values ($\geq$5) the ellipsoidal
modulation is largely insensitive to $q$ and hence can provide
excellent constraints on $i$. For full details of the light curve
modelling see Tjemkes et al (1986), Orosz \& Bailyn (1997) and Shahbaz
et al (2003), the results from which are in table~{\ref{tab:SXTs}}.

\subsection{Mass Determinations and Limitations}

To derive full XRN orbital solutions requires assuming that the
quiescent secondary fills its Roche lobe (reasonable as doppler
tomography of quiescent XRN reveals ongoing mass transfer; Marsh et al
1994), and that the light curve is {\it not} contaminated by any other
emitting region.  This latter assumption is questionable, since
emission lines are definitely attributable to a quiescent disc, but
residual X-ray heating and coronal activity on the secondary might
also be present (e.g. Bildsten \& Rutledge 2000).  For this reason, secondary light
curves are usually derived in the IR whenever possible.  The disc
contribution in the optical around H$\alpha$ is obtained as a
by-product of the spectral type determination, and is typically
$\leq$10\%.  It should therefore be even less in the IR for a typical
disc spectrum. However, the outer disc can be an IR emitter in CVs
(Berriman et al 1985) and hence contaminate the light curves (Sanwal
et al 1996).  This is potentially significant, since a contaminating
(and presumably steady) contribution will reduce the amplitude of the
ellipsoidal modulation, and hence a lower $i$ would be inferred,
leading to an erroneously high mass for the compact object.

Nevertheless Shahbaz et al (1996, 1999a)
showed via IR K-band spectroscopy of V404 Cyg and A0620-00 that any
contamination must be small and hence the masses derived need (at most)
to be reduced by only small amounts.  Furthermore, in studying the
non-orbital optical variability in V404 Cyg, Pavlenko et al
(1996) found that (as first noted by Wagner et al
1992) the ellipsoidal modulation could be discerned
underlying the substantial (short-term) flickering in the light curve.
Interpreting the flickering as a completely independent component
(recently verified by Hynes et al 2002a), Pavlenko et al
showed that the {\it lower envelope} of this light curve (rather than
the mean) produced an ellipsoidal light curve which, when fitted as
described above, gave essentially identical results to those obtained
from the IR ellipsoidal fitting, thereby providing further weight to
the significance of the final mass determinations in
table~{\ref{tab:SXTs}}.  Note that the values for Cen X-4 and
XTE~J2123-058 (both neutron star XRN, identified on the basis of their
type I X-ray bursts) have been derived exactly by the method outlined
here and yield values in excellent accord with those expected for a
neutron star.  Table~{\ref{tab:SXTs}} contains only those systems for
which a dynamical study of the companion star has been performed (all
via direct detection of the spectral signature of the mass donor,
except for GX339-4 - see section~{\ref{sect:LumLMXBs}}).  There are a
number of additional BHC (based on their X-ray and other properties)
for which such studies have not yet been possible (secondary too faint or
too heavily reddened), and these are listed in chapter 4.

\begin{figure}[t]
\begin{center}
\begin{picture}(100,250)(50,30)
\put(0,0){\includegraphics{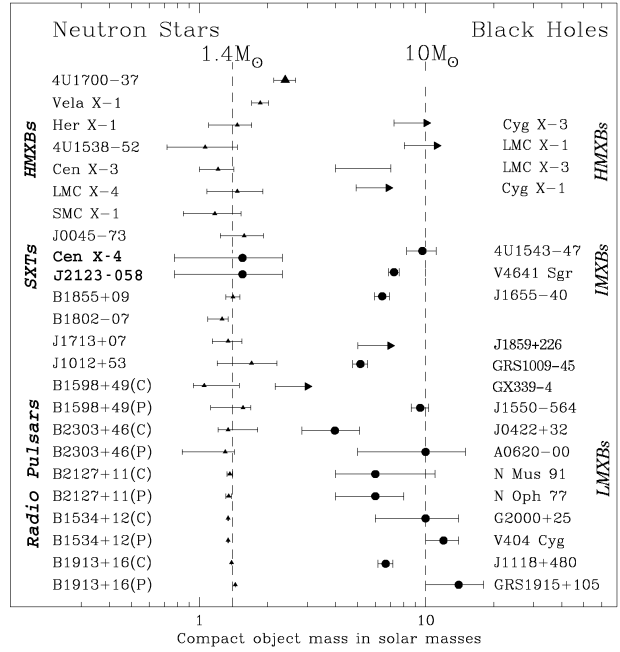}}
\noindent
\end{picture}
\caption {The mass distribution of neutron stars and black 
holes.  Note the
remarkably narrow spread of NS masses, and the large factor by
which the BH masses exceed the (canonical) maximum NS mass of 3.2$M_\odot$.
\label{masses}}
\end{center}
\end{figure}

For completeness, we have also included in table~{\ref{tab:SXTs}} the
dynamically determined HMXB compact object mass determinations.  In
spite of an extremely accurately determined $f(M)$, there are still
large and systematic uncertainties in the mass of Cyg X-1's supergiant
primary HDE 226868 and the difficulty in constraining the binary
inclination (it does not eclipse).  This has most recently been
addressed by Herrero et al (1995) whose high resolution optical
spectroscopy and detailed atmospheric modelling give a mass range of
12--19~M$_\odot$ for the OBI primary.  With $i$ relatively poorly
constrained to 28--67~degrees, the compact object mass must be in the
range 4--15~M$_\odot$.  HDE 226868 has also been studied by LaSala et
al (1998) and Brocksopp et al (1999) who have derived a much improved
orbital ephemeris, and Ba\l uci\'nska-Church et al (2000), who used
this new ephemeris to study the distribution of X-ray dips as a
function of orbital phase (there is a strong orbital modulation).
In spite of this large uncertainty in compact object mass (a feature
shared by the LMC HMXBs LMC X-1 and LMC X-3), Cyg X-1 remains a very
strong BHC, and its X-ray properties are still invoked as black hole
characteristics (see chapter 4).

Interestingly most of the compact object masses so far determined lie
in the range 6--14M$_\odot$, so that even the lowest value is well
above the theoretical neutron star maximum mass of $\sim$3.2M$_\odot$
(and even further from the directly measured neutron star masses of
1.35M$_\odot$; see Thorsett \& Chakrabarty 1999 and
fig~{\ref{masses}}).  There is only one compact object mass currently
in the range 2--5M$_\odot$ (the recently determined value for
GRO~J0422+32 by Gelino \& Harrison 2003, but GRS1009-45 is close),
which is curious and may be a selection effect.  Certainly V518 Per
and MM Vel are two of the faintest in quiescence, but it may also be
possible that transient behaviour is suppressed.  In such cases mass
determination becomes difficult as the disc dominates the optical
light, but alternative methods are now possible in certain
circumstances (see section~{\ref{sect:LumLMXBs}}), or if the donor is
sufficiently bright to be detectable even against the glare of the
disc (GRS~1915+105, Cyg X-2).

\subsection{Abundance Analyses of the Mass Donors}

The absence of (significant) X-ray heating allows for detailed
chemical analyses of the mass donors in the brighter of the quiescent
XRN.  An immediate (and unexpected) by-product of the high resolution
radial velocity study of V404 Cyg was the discovery of strong LiI
$\lambda$6707 absorption (Mart\'\i{}n et al 1992).  Li is
typically present in young, pre-main sequence and T Tau stars, but is
destroyed by subsequent convection in late-type stars, and normal (solar)
abundances are a thousand times lower.  Similar Li
enhancements are found in many (but not all) quiescent XRN (see
Mart\'\i{}n et al 1996), including the NS XRN Cen X-4, so
this is {\it not} a potential BH signature.

As highly evolved objects which are extremely unlikely to have
retained such high Li abundances, mechanisms have been sought that
create Li within XRN, which could be important for the wider study of
the galactic Li enrichment relative to the halo.  Li has been seen in
XRN with a wide range of $P$ (and hence donor sizes), but their common
feature is the recurrent, high $L_X$ X-ray outbursts (CVs do {\it not}
exhibit Li, Mart\'\i{}n et al 1995), which led Mart\'\i{}n et al
(1994) to suggest that spallation could produce Li in large quantities
close to the compact object, and might explain the 476keV $\gamma$-ray
feature observed during the N Mus 1991 outburst (Sunyaev et al 1992;
Chen et al 1993).  Originally interpreted as the gravitationally
redshifted $e^-$-$e^+$ 511keV line (see also Kaiser \& Hannikainen
2002), it might instead be associated with the $^7$Li 478keV line.
Some of this Li would be transferred to the secondary during
subsequent large mass outflows.  However, Bildsten \& Rutledge (2000)
point out that the XRN Li abundances are only slightly higher than
those detected in the chromospherically active RS CVn systems and also
the pre-CV binary V471 Tau.  This suggest a possible link with coronal
activity, although it has been argued by Lasota (2000) that this
cannot account for the levels of X-ray emission seen in quiescent XRN.

More dramatic processes, and a possible link of XRN with hypernovae
has resulted from the discovery by Israelian et al (1999)
of substantial (factors 6 to 10) increases in the $\alpha$-element
abundances (O, Mg, Si, S) in very high resolution spectra of the F
sub-giant donor in GRO~J1655-40.  Other elements (such
as Fe) show entirely normal (solar) abundances, and so Israelian et al
infer that the companion star was bathed in these elements which were
created by explosive nucleosynthesis during the supernova that formed
the compact object.  Interestingly, Israelian et al point out that the
current BH mass estimate ($>$5.5\msun;
table~{\ref{tab:SXTs}}) requires a hypernova explosion in order to
account for the observed abundances (see also Podsiadlowski et al
2002).

\subsubsection{UV Spectroscopy}

\begin{figure*}
\begin{center}
\psfig{angle=90,height=2.0in,width=5in,file=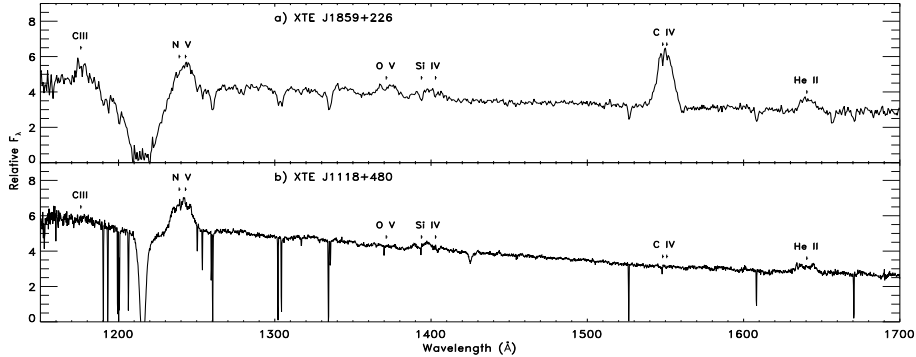}
\caption{HST/STIS UV spectra of XTE J1859+226 (upper), which
shows normal line ratios (CIV more prominent than NV), whereas
XTE J1118+480 (lower) has CIV and OV absent, yet NV is prominent (from Haswell et al 2002).
}
\label{1118uvspec}
\end{center}
\end{figure*}

The evolutionary history of XRN can also be inferred from relative
abundances of nuclear processed material, e.g.  through UV
spectroscopy.  A comparison by Haswell et al (2002) of J1118+480 and
J1859+226 (both observed by HST/STIS near outburst peak) revealed
remarkably similar strengths of NV and SiIV, yet J1118+480 exhibited
no CIV or OV (fig~{\ref{1118uvspec}}).  This is very surprising since
CIV and NV are both resonance lines of Li-like ions and therefore
produced under similar physical conditions.  This dramatic difference
between these two short period systems implies that the mass donor in
J1118+480 must have lost its outer layers, thereby exposing inner
material which has been mixed with CNO-processed matter from the core
(the evolution of the C/N ratio as a function of $M_2$ is shown in
chapter 13 and Ergma \& Sarna 2001).  Such an interpretation requires
that mass transfer began at an initial secondary mass of
$\sim$1.5\msun and $P$ of $\geq$12h, but this raises interesting
questions as to how the binary then evolves to its current $P=$4.1h
(see chapter XX).  However, this does support the results of Smith \&
Dhillon (1998) whose detailed analysis of the secondary properties in
CVs shows that (up to $P\sim$7-8h) they are indistinguishable from
main-sequence stars in detached binaries, whereas the XRN secondaries
have much larger radii and are therefore evolved.

The HST spectra of J1859+226 (Hynes et al 2002b) also
display a strong $\lambda$2175 feature that allows for an accurate
determination of E(B-V).  Combined with optical spectroscopy, Hynes et
al find that the UV/optical region can be well fit by a standard, but
{\it irradiated} disc.  As the outburst progresses, the irradiated
component declines and the implied outer disc radius reduces, implying
that the system fading is a result of a {\it cooling wave} moving
inwards (see also Lasota 2001).

\subsection{Outburst/Decline Properties}

\subsubsection{Light curve shapes and average properties}

The wide variety of X-ray light curve shapes and outburst behaviour
has been reviewed by Tanaka \& Shibazaki (1996), Chen et
al (1997) and updated in chapter 4.  But the basic shape
of a fast rise and exponential decay (fig~{\ref{a0620}) has
been interpreted as a disc instability (see Cannizzo
1998, King \& Ritter 1998 and chapter 16) and
is a natural consequence of an X-ray irradiated disc being maintained
in a hot (viscous) state (and hence producing much longer outbursts
than in dwarf novae, see King \& Ritter).  Furthermore, Shahbaz et al
(1998) have shown that SXTs can exhibit both
exponential and linear decays, the latter occurring if the SXT
outburst $L_X$ is insufficient to ionise the disc outer edge
(see fig~{\ref{Lcrit}} for XRN with known $P_{orb}$ and
well-defined light curve shape).

\begin{figure}[t]
\begin{center}
\begin{picture}(100,250)(50,30)
\put(0,0){\includegraphics{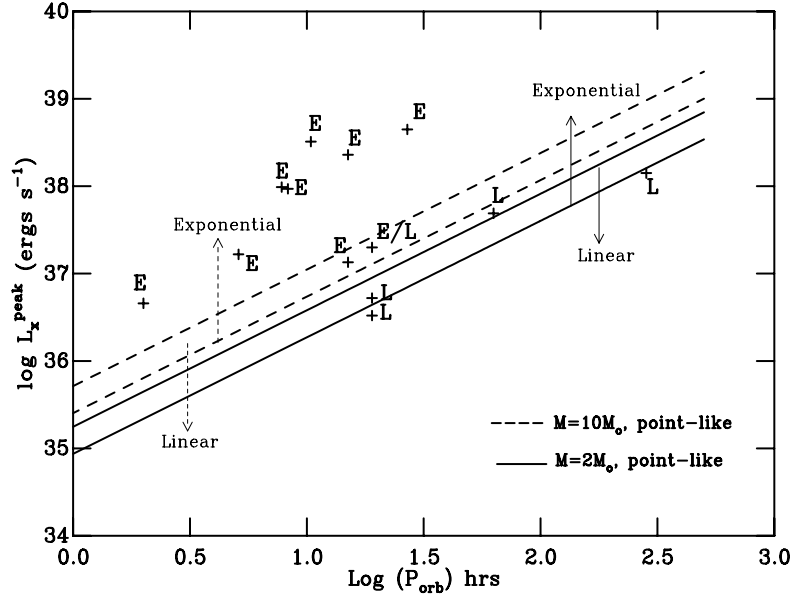}}
\noindent
\end{picture}
\caption {The critical $L_X$ needed to ionise the entire
accretion disc (Shahbaz et al
1998), for total masses of 2 and
10$M_\odot$ corresponding to NS and BH SXTs
respectively.  These $L_X$ are a factor 2 smaller for
exponential (E) decays compared to linear (L) ones, due to the
difference in the disc circularisation and tidal radii.  The
SXTs shown are SAX J1808.4-3658, GRO J0422+32, A0620-00, GS2000+25,
GS1124-68, Cen X-4, Aql X-1, 4U1543-47, GRO J1655-40 and GRO J1744-28.
\label{Lcrit}}
\end{center}
\end{figure}

Additionally, Shahbaz \& Kuulkers (1998)~\cite{sk98} have shown
(fig~{\ref{amplitude}}) that the optical outburst amplitude ${\Delta}V$
is related to $P_{orb}$ (if $P_{orb}<$1 day) according to:

\begin{equation}
{\Delta}V = 14.36 - 7.63 \log{P_{orb}(hrs)} 
\end{equation}

which is essentially due to longer $P_{orb}$ systems
having larger and hence brighter mass donors (since they must be
Roche-lobe filling).  Hence a guide to $P_{orb}$ can be immediately
obtained once $\Delta$V is known.

\begin{figure}[t]
\begin{center}
\begin{picture}(100,250)(50,30)
\put(0,0){\includegraphics{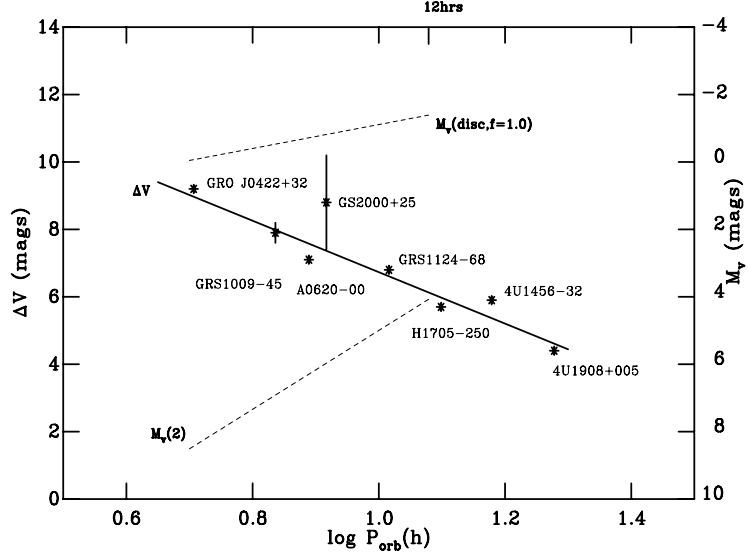}}
\noindent
\end{picture}
\caption {Visual outburst amplitude of SXTs as a function of $P_{orb}$
(Shahbaz \& Kuulkers 1998).  The dashed lines show
the calculated disc $M_V$ in an outbursting SXT and the donor $M_V$
according to Warner (1995).
\label{amplitude}}
\end{center}
\end{figure}

\subsubsection{Do Superhumps Occur in LMXBs? \label{sect:shumps}}

The high $q$ values of XRN are also implied by the detection of {\it
superhumps} in their optical light curves (during outburst decay;
O'Donoghue \& Charles 1996).  As discussed in Warner (1995 and
references therein), superhumps occur during {\it superoutbursts} of
SU UMa-systems and are attributed to tidal stressing of the accretion
discs in high $q$ interacting binaries.

A wide range of models were proposed to account for this remarkable
feature, but it is now widely accepted that SU UMa superhump are due
to tidal energy release in an elliptical, precessing disc which occurs
when the disc expands beyond its stability radius (Whitehurst \& King
1991).  However, this requires a high $q$ ($\geq$3) or else the
stability radius is outside the tidal radius, and hence matter is
simply lost from the system.  With white dwarf masses typically
$\leq$1M$_\odot$ this implies very low mass secondaries and is
expected to occur in short $P$ systems (almost all SU UMas are below
the period gap).  Mineshige et al (1992) showed that the period
excess $\Delta$ (=$(P_{SH}-P_{orb})/P_{orb}$) is related to $q$ via

\begin{equation}
\Delta = \frac{3}{4} q^{-1/2} (1 + q)^{-1/2} (\frac{R_D}{a})^{3/2}
\end{equation}

where $R_D$ is the disc radius and $a$ the orbital separation, and
hence there is the potential for deriving some dynamical information
from a simple measurement of the superhump $\Delta$.

This basic model for the superhump phenomenon makes no requirement on
the nature of the compact object, merely the value of $q$, implying
that high $q$ LMXBs, in which the compact object is now a NS or BH,
should also display this effect.  However, the much greater depth of
the compact object potential well ($\geq$10$^3$) in LMXBs implies that
their intrinsic $L_X$ exceeds that of CVs by the same factor, and
hence that X-ray irradiation of the disc produces optical emission
that far exceeds the {\it intrinsic} light of the disc, i.e. the tidal
forces that produce superhumps should be swamped by irradiation.
Nevertheless, observations have revealed effects related to the high
$q$ in both XRN and double-degenerate LMXBs (such as X1626-671 and
X1916-053, see section~{\ref{sect:lps}}).  While the latter are
confirmed (via X-ray bursts) as NS, their ultra-short $P$ (41 and 50
mins respectively) require a degenerate, and hence very low mass
donor, so that $q$ is very high in both cases.

This potentially important effect has been explained by Haswell et al
(2001) who utilised the results of disc simulations in
extreme $q$ binaries by Murray (2000).  These
simulations clearly demonstrated the onset of the 3:1 instability that
leads to the precessing, elliptical disc (see Whitehurst \& King 1991), but they
also showed that the effective {\it area} of the disc is also
modulated on the superhump period.  And if the disc area is modulated
at a given period, then so will the reprocessed light, which is the
dominant factor in LMXBs. This effect is also independent of $i$,
as observed, unlike reprocessing in the heated face of the companion.
The significance of this feature in transient LMXBs where the compact
object is suspected of being a black hole is that its observation can
lead to an estimate of the compact object mass even while the outburst
is ongoing.  However, there are many ``steady'' LMXBs (those in bold
in table~{\ref{tab:LMXBs}}) where the periodicity observed has not yet
been confirmed as orbital in origin (and requires either eclipses or a
radial velocity curve).  Such systems are ideal targets for future
application of the Bowen fluorescence mechanism
(section~{\ref{sect:LumLMXBs}}).

\begin{table*}
\begin{center}
\caption{\label{tab:LMXBs}LMXB Modulation Properties}
{\small
\begin{tabular}{lcll} 
\hline
{\em Source} &  Period & Nature of &
X-ray type \\
 & (hrs) & modulation &\\ 
\hline

X1820-303 & {\bf 0.19} & X-ray & Burster, glob.cl. \\
4U~1850--087   & {\bf 0.34}  & UV &Burster, glob.cl. \\
X1626-673 & {\bf 0.7}  & opt sideband & Burster, Pulsar \\
X1832-330 & {\bf 0.73} & UV &Burster, glob.cl. \\
X1916-053 & {\bf 0.83} & X-ray, opt & Burster, Dipper \\
J1808.4-3658 & 2.0 & pulsation RV & Burster, Pulsar, Transient \\
X1323-619 & {\bf 2.9}  & X-ray dip & Burster, Dipper \\
X1636-536 & {\bf 3.8} & opt & Burster\\
X0748-676 & 3.8 & eclipsing & Burster, Dipper, Transient \\
X1254-690 & {\bf 3.9} & X-ray dip & Burster, Dipper \\
X1728-169 & {\bf 4.2} & opt &        \\
X1755-338 & {\bf 4.4} & X-ray dip & Dipper  \\
X1735-444 & {\bf 4.6} & opt & Burster \\
J0422+32 & 5.1 & opt RV & BH,  Transient \\
X2129+470 & {\bf 5.2} & opt & ADC     \\
X1822-371 & 5.6 & eclipsing & ADC     \\
J2123-058 & 6.0 & eclipsing & Burster, Transient \\
N Vel 93 & 6.9 & opt RV & BH,  Transient \\
X1658-298 & {\bf 7.2} & X-ray dip & Burster, Dipper \\
A0620-00 & 7.8 & opt RV & BH,  Transient \\
G2000+25 & 8.3 & opt RV & BH,  Transient \\
A1742-289 & 8.4 & eclipsing & Burster, Transient \\
X1957+115 & {\bf 9.3} & opt &         \\
N Mus 91 & 10.4 & opt RV & BH,  Transient \\
N Oph 77 & 12.5 & opt RV & BH,  Transient \\
Cen X-4 & 15.1 & opt RV & Burster, Transient \\
X2127+119 & 17.1 & eclipsing & Burster, ADC, glob.cl. \\
Aql X-1 & {\bf 19} & opt & Burster, Transient \\
Sco X-1   & {\bf 19.2} & opt & Prototype LMXB \\
X1624-490 & {\bf 21} & X-ray dip & Dipper \\
N Sco 94 & 62.6 & opt RV & BH,  Transient \\
V404 Cyg & 155.4 & opt RV & BH,  Transient \\
2S0921-630 & 216 & eclipsing & ADC \\
Cyg X-2 & 235 & opt RV & Burster \\
J1744-28  & 283   & pulsation RV & Burster, Pulsar, Transient \\

\hline
\end{tabular}
}
{\footnotesize adapted from Charles (2001), van Paradijs \& McClintock (1995) and van Paradijs (1998)}
\end{center}
\end{table*}

\subsection {Outburst spectroscopy of GRO~J1655-40} 

Obtaining dynamical information during the outburst/decline phases
would be extremely valuable, as the majority of quiescent SXTs are
extremely faint.  The superluminal transient GRO~J1655-40 (see Mirabel
\& Rodriguez 1999) exhibited extended intervals of activity in 1994
and 1996, during which Soria et al (1998) followed the orbital
behaviour of its strong, complex H$\alpha$ emission, but the
asymmetric profile precluded its use as a dynamical tracer of the
compact object.  However, the broad wings of the double-peaked
HeII$\lambda$4686 emission were found to be representative of the
inner disc regions, giving a radial velocity curve in anti-phase with
the companion star to within 9$\pm$20$^o$.  Unfortunately, it is still
difficult to use this for dynamical information as the HeII mean
velocity is -182 km~s$^{-1}$, whereas the secondary absorption lines
give -142 km~s$^{-1}$ (both with very small errors).  This systematic
blue shift in the disc (emission) lines suggests the presence of a
substantial disc wind during X-ray high states.

Combined with one of the earliest spectral types (mid-F), the high
$\gamma$-velocity suggests that GRO~J1655-40 might have been formed
via accretion-induced collapse of a neutron star (Brandt et al
1995).  Furthermore, after its initial 1994 outburst, a
subsequent optical rebrightening began $\sim$6 days before the X-rays,
indicating an ``outside-in'' outburst of the accretion disc (Orosz et
al 1997).  This substantial delay is due to the inner
disc needing to be re-filled before accretion onto the compact object
begins again, it having been evaporated by hard X-rays from the ADAF
flow during quiescence (see Hameury et al 1997).

GRO~J1655-40 is extremely important amongst the XRN because of its
brightness in quiescence, thereby yielding high quality photometric
light curves and high resolution phase-resolved spectroscopy, from
which the orbital system parameters can be derived (Orosz \& Bailyn
1997; but note the error analysis of van der Hooft et al 1997).  The
early spectral type also means that GRO~J1655-40 has a low mass ratio
($q{\sim}3$), and hence the ellipsoidal modulation is sensitive to
both $q$ and $i$ (the latter being constrained by the observed grazing
eclipse, see fig~{\ref{J1655lc}}).  The advanced evolutionary state of
the secondary (which has been addressed by Kolb et al (1997), Kolb
(1998) and Beer \& Podsiadlowski (2002) is driving the much higher
mass transfer rate, but it returns to the transient domain following
temporary drops in \.{M}.

\subsection {Effects of Irradiation}

{\em (a) Echo Mapping.}  Many of the X-ray transients are sufficiently
bright during outburst, that rapid X-ray variability provides a
signature that allows ``echo mapping'' to be employed, which can probe
the binary geometry.  Although well-developed for mapping AGN
structure (see e.g. Horne 1999a), its application to
LMXBs requires high time resolution ($\leq$ secs) at optical/UV
wavelengths, because of the $\sim$seconds light-travel time across
short-period LMXBs.  This imposes a severe constraint on current
technology, and is provided by few observatories due to
the limitations of normal CCD cameras.  However, HST does
provide this facility, and Hynes et al (1998) obtained
simultaneous X-ray (RXTE) high speed optical/UV FOS spectroscopy of
GRO~J1655-40 during its extended outburst, that showed the optical
lagging the X-ray by $\sim$10--20s at times of X-ray flaring activity
(see fig~{\ref{echoes}}).  With a relatively long $P_{orb}$
(2.6d) and well-established orbital ephemeris, this optical delay can
be explained entirely by reprocessing within the accretion disc and
not heating of the secondary.

More recently, Kanbach et al (2001) and Hynes et al
(2003a) performed a similar study (using fast ground-based
photometry and HST/STIS UV spectroscopy respectively) on
XTE~J1118+480.  While the absolute HST timing is still uncertain,
these data do show the longer UV wavelengths ($\sim$2700\AA) lagging
the shorter ($\sim$1400\AA) by about 0.25s.  However, the scale of
variability suggests that the bulk of this component is due to
synchrotron emission, and not the accretion disc.

The long-term goal of these studies is to follow such behaviour
throughout several orbital cycles, which in principle would allow the
system geometry to be fully mapped.  Unfortunately, this is a very
difficult observational programme, particularly given that at times of
X-ray outburst the binary orbital ephemeris is rarely known.  But
given the typical recurrence timescales of XRN of 10-50 years, such
opportunities will grow in future and echo mapping could become an
important technique.

\begin{figure}[t]
\begin{center}
\includegraphics[angle=90,scale=0.40]{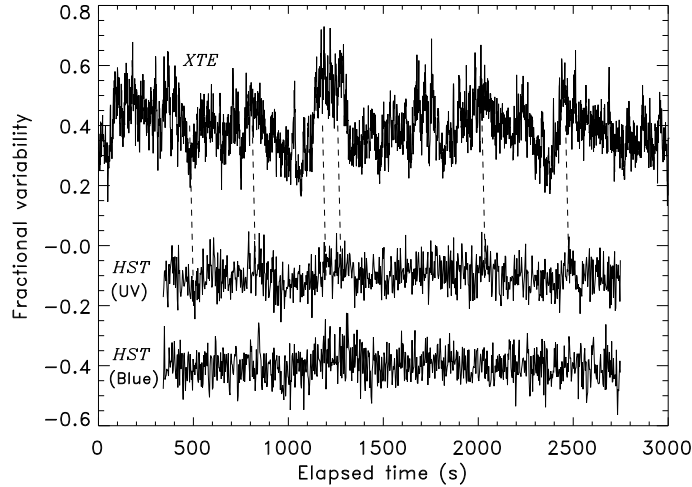}
\caption {X-ray (RXTE) and optical/UV (HST)
lightcurves of GRO~J1655-40 obtained in 1996 by Hynes et al (1998) which
shows the $\sim$10--20s delay of the optical response to the X-ray
flares, and are connected by dashed lines.
\label{echoes}}
\end{center}
\end{figure}

{\em (b) Distortion of the Radial Velocity Curve.}  With the X-ray
variability signature clearly visible in the disc's optical/UV
emission, it is likely that the (very high) outburst $L_X$ has a
significant impact on the secondary's atmosphere.  The
majority of XRN have donors that, in any case, are completely
undetectable during outburst, but the small sub-group of intermediate
mass donors (see table~{\ref{tab:SXTs}}) are sufficiently luminous to
be easily spectroscopically visible at all times.  This is the case
with GRO~J1655-40 and the original dynamical study (Orosz \& Bailyn
1997) should, ideally, have been undertaken during X-ray
quiescence.  But in order to obtain full orbital phase coverage they
did use some spectra taken during outburst.  Under such circumstances,
care must be taken to ensure that irradiation of the secondary
has not systematically distorted the radial velocity curve.

Phillips et al (1999) tested for the presence of this
effect by fitting an elliptical orbit to the Orosz \& Bailyn data,
which gave a highly significant $e$=0.12$\pm$0.02.  With the ratio of
incident X-ray to local stellar flux at the donor surface of
$\sim$7, it is not surprising that such effects can be important.
Fitting an irradiated model of the secondary with free parameters $q$,
$i$ and disc opening angle (which shadows the secondary's equatorial
regions) Phillips et al obtained $q$=2.8 and $M_X$ in the range
4.1--6.6M$_\odot$.  A completely quiescent radial velocity curve was
subsequently obtained by Shahbaz et al (1999, 2000) which
confirmed this analysis by constraining $q$ to the range 2.29--2.97
and $M_X$ to 5.5--7.9M$_\odot$, and updated further by Shahbaz
(2003) to the values given in table~{\ref{tab:SXTs}}.
Hence, it is still possible (Brandt et al 1995) that the high
$\gamma$-velocity of GRO~J1655-40 might indicate the formation of a
relatively low mass BH via accretion-induced collapse of a neutron
star.

\subsection{Spectroscopy of Luminous LMXBs \label{sect:LumLMXBs}}

\subsubsection{Sco X-1 and X-ray irradiation of the companion}

The detailed dynamical analysis of the XRN in
section~{\ref{sect:qXRN}} is only possible because of their very long
quiescent intervals, during which the optical emission is dominated by
the companion.  This accounts for the remarkable fact that XRN account
for the majority of detailed LMXB mass constraints
(table~{\ref{tab:SXTs}}).  And since few LMXBs are X-ray pulsars,
remarkably little is known of the fundamental parameters of the X-ray
luminous (``steady'') LMXBs, as their companion stars are perpetually
hidden from view in the glare of the X-ray illuminated accretion disc.
Indeed, in spite of being the brightest LMXB in the sky (both
optically and in X-rays), key parameters of Sco X-1 have only been
determined quite recently.  VLBA observations of its twin radio lobes
give $d=$2.8$\pm$0.3~kpc (Bradshaw et al 1999) and $i=$44$\pm$6
degrees (Fomalont et al 2001). However, the donor has not been
detected spectroscopically in either the optical (Schachter et al
1989) or IR (Bandyopadhyay et al 1997, 1999), and the optical spectrum
is dominated by the X-ray irradiated disc.  This situation has been
transformed by Steeghs \& Casares (2002) who obtained the first high
resolution optical spectroscopy that revealed very sharp emission
features from the Bowen $\lambda\lambda$4640-4650 emission blend
(fig~{\ref{scox1spec}}).

\begin{figure}
\includegraphics[width=9cm,angle=0]{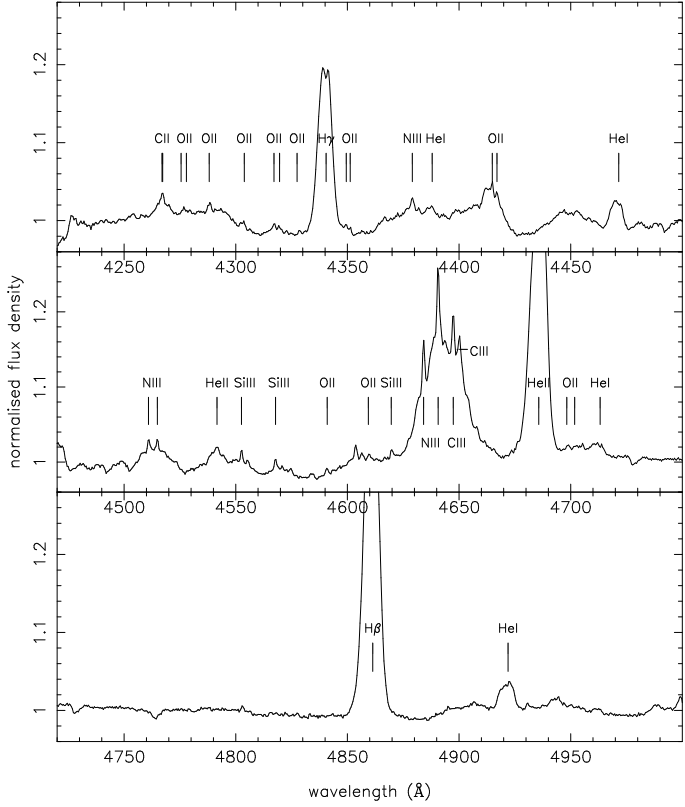}
\caption{
High resolution blue spectrum of Sco X-1 (Steeghs \& Casares 2002) which demonstrates the complex emission profile of the $\lambda\lambda$4640-50 Bowen blend.  Note particularly the very sharp components of NIII and CIII.
}\label{scox1spec}
\end{figure}

These broad emission components produce a radial velocity curve
(LaSala \& Thorstensen 1985) whose phasing with respect to the
lightcurve (Gottlieb et al 1975) confirms their production in the
disc.  However, the sharp components move in {\it anti-phase} to the
broad emission and are therefore associated with the companion's {\it
heated} face (fig~{\ref{scox1rv}}).  Note that the Bowen motion is not
strictly sinusoidal, as the irradiation dominates on the inner Roche
lobe that faces the X-ray source (although there will also be some
disc shadowing which will ameliorate this), and hence only the limit
$K_2>$77~\kms is obtained.  However, doppler tomography (see Marsh
2001 for a detailed review of this technique) of Sco X-1 with these
data (fig~{\ref{scox1dop}}) shows strong emission from the secondary
and indicates $K_2>$87~\kms.  There is a slight offset with the
anti-phased HeII emission ($\Delta\phi\simeq$0.1; likely due to hot
spot contamination as in CVs, see Warner 1995), but the amplitude
implies $K_1<$53~\kms.  The doppler tomograms show that there is a
HeII component on the secondary as well (but not HeI or Balmer
emission, these arise completely on the disc), and this distorts the
disc radial velocity curve.  Searching only for the symmetric
component gives $K_1\simeq$40~\kms, and hence $q>$2.2.  However,
Steeghs \& Casares argue that the likely offset of the heated face
combined with the radio-determined $i$ yield $M_2\sim$0.4\msun if the
neutron star has a canonical 1.4\msun.  This requires an evolved
secondary and hence \mdot$\sim$8$\times$10$^{-10}$\msun~y$^{-1}$ (King
et al 1996), which gives a disc $T>$6500K which is stable.

\begin{figure}
\centering
\begin{minipage}[c]{0.5\textwidth}
\centering
\includegraphics[width=5cm]{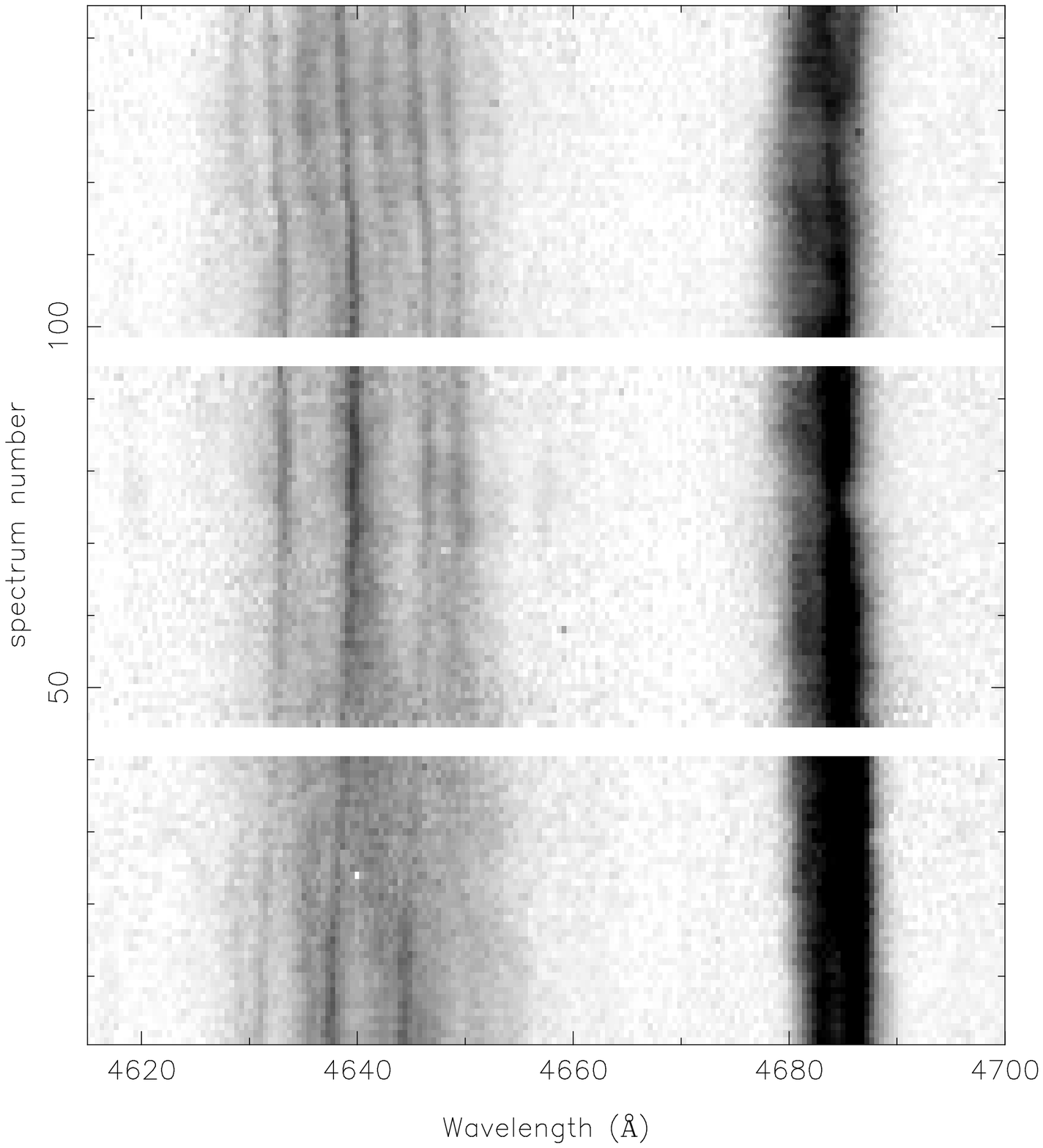}
\end{minipage}%
\begin{minipage}[c]{0.5\textwidth}
\centering
\includegraphics[width=5cm,angle=-90]{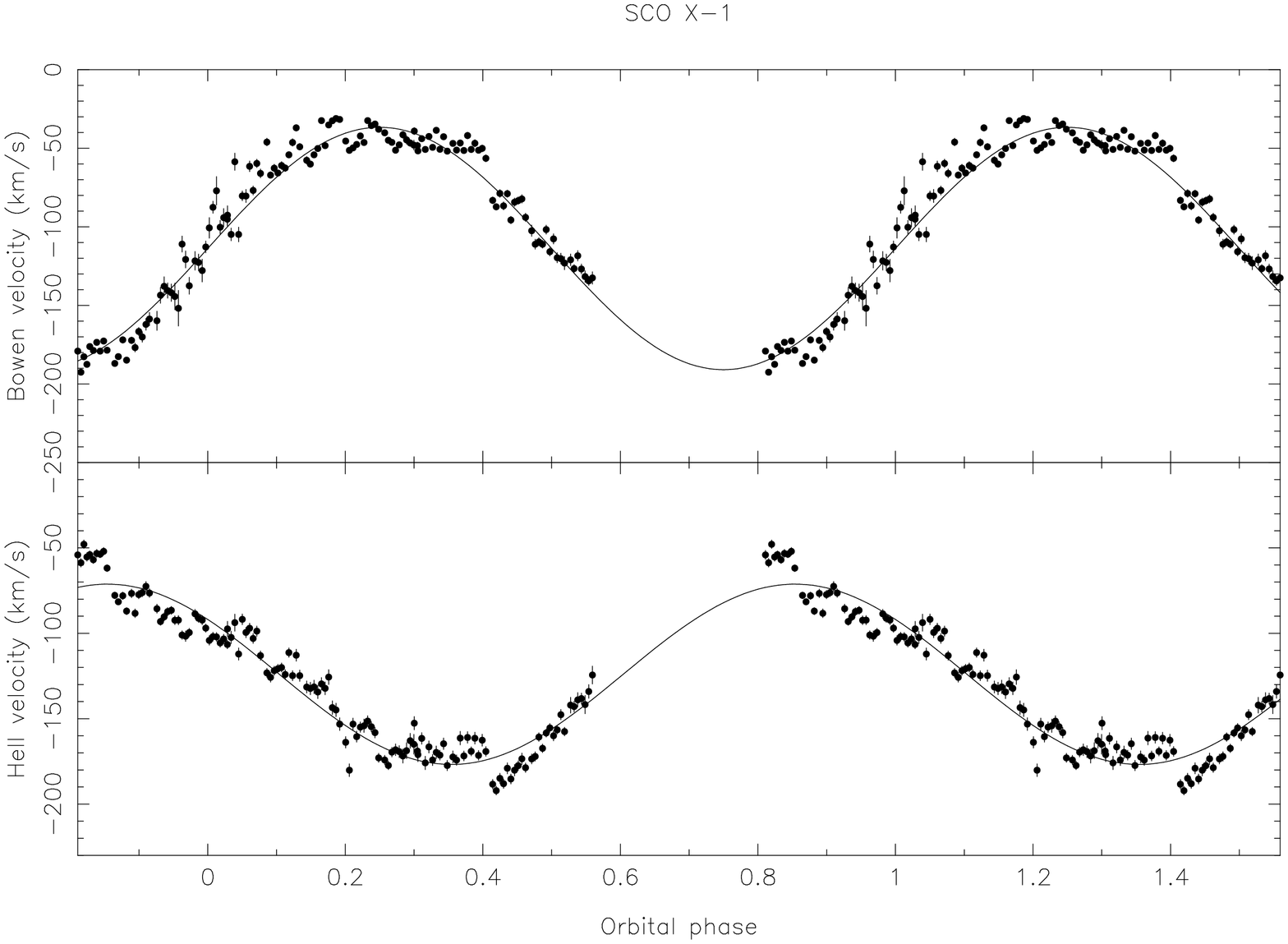} 
\end{minipage}
\caption{Trailed spectrogram ({\it left}) and radial velocity curves ({\it right}) of the HeII and Bowen emission from Sco X-1 (Steeghs \& Casares 2002).
\label{scox1rv}}
\end{figure}

\begin{figure}
\includegraphics[width=8cm,angle=0]{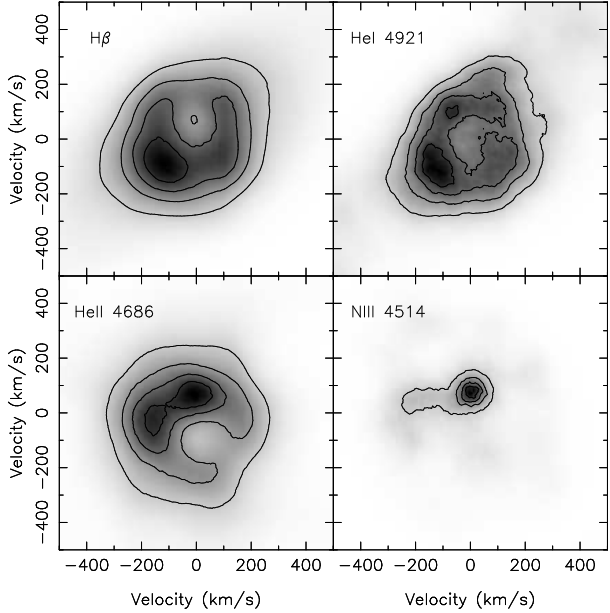}
\caption{
Doppler tomograms of Sco X-1 (Steeghs \& Casares 2002) where the origin is the binary centre of mass.  Note the strong NIII emission located on the companion, whereas the other emission lines' ring-like structure is typical of accretion discs (although HeII also has a component on the companion).
}\label{scox1dop}
\end{figure}

\subsubsection{GX339-4 and XRN in outburst}

This powerful new technique has already been applied to the (almost)
steady LMXB BHC GX339-4 (Hynes et al 2003b).  First
proposed as a BHC (Samimi et al 1979) based on its Cyg X-1-like
X-ray fast variability, GX339-4 moves regularly through high, low and
``off'' states (ref) on timescales of a year or two.  Associated with
these, the optical counterpart (V821 Ara) varies between V$<$15 at
peak to $>$20 at minimum.  Yet even at its faintest, there has been no
detection of the (presumed) cool mass donor (Shahbaz et al 2001).
Consequently, even $P_{orb}$ has been ill-constrained, and so
the return to an X-ray bright phase in mid-2002 provided an
opportunity to apply the Bowen fluorescence technique.

The Bowen blend is indeed strong when the source is X-ray bright, and
smoothly moving sharp features were detected which suggest $P=$1.76d,
compared to the (non-confirmed) 14.8h period of a decade earlier
(Callanan et al 1992; see also Cowley et al 2002).  The resulting
$K_2$ velocity leads to the $f(M)$ given in table~{\ref{tab:SXTs}},
finally confirming its BHC status.  What is curious in this case is
the fact that the sharp features are not always visible (Hynes et
al 2003b), suggesting that there are significant
intervals during which the face of the donor is effectively shielded
from direct illumination.  This could arise if the disc is either
warped or is elliptical and precesses in a way that it approaches the
donor (refs).  Nevertheless, these results demonstrate the potential
power of this technique and we anticipate substantial progress in the
coming decade in determining, for the first time, the dynamical
properties of the luminous LMXBs.

\subsection{Orbital light curves and tomography of LMXBs}

\subsubsection{XTE~J2123-058}

The 1998 NS XRN J2123-058 is an ideal system with which to demonstrate
the range of LMXB optical light curves.  It has a
relatively short (6h) period and orbital light curves were obtained
from outburst peak through decline and into quiescence by Zurita et al
(2000); see also Soria et al 1999). The high $i$
produces a spectacular modulation that evolves from
triangular to double-humped (fig~{\ref{2123lcs}}) as the relative
contributions of the disc, X-ray heated companion and
ellipsoidal components change during the decline.  These exquisite
light-curves were modelled by Zurita et al (see also the re-analysis
by Shahbaz et al 2003) to constrain 
$i$ to be 73$\pm$4$^\circ$. J2123-058 also produced a value of
$\xi$=$B_0+2.5{\log}F_X({\mu}Jy)$=21.9, which compares remarkably well
with the canonical value for LMXB X-ray heating of 21.8$\pm$1 from van
Paradijs \& McClintock (1994).  The observation of an X-ray burst also
requires the neutron star to be directly visible, and hence provides a
tight constraint on the disc flare angle ($\alpha<$90-$i$).

\begin{figure}
\centering
\begin{minipage}[c]{0.5\textwidth}
\centering
\includegraphics[width=5cm]{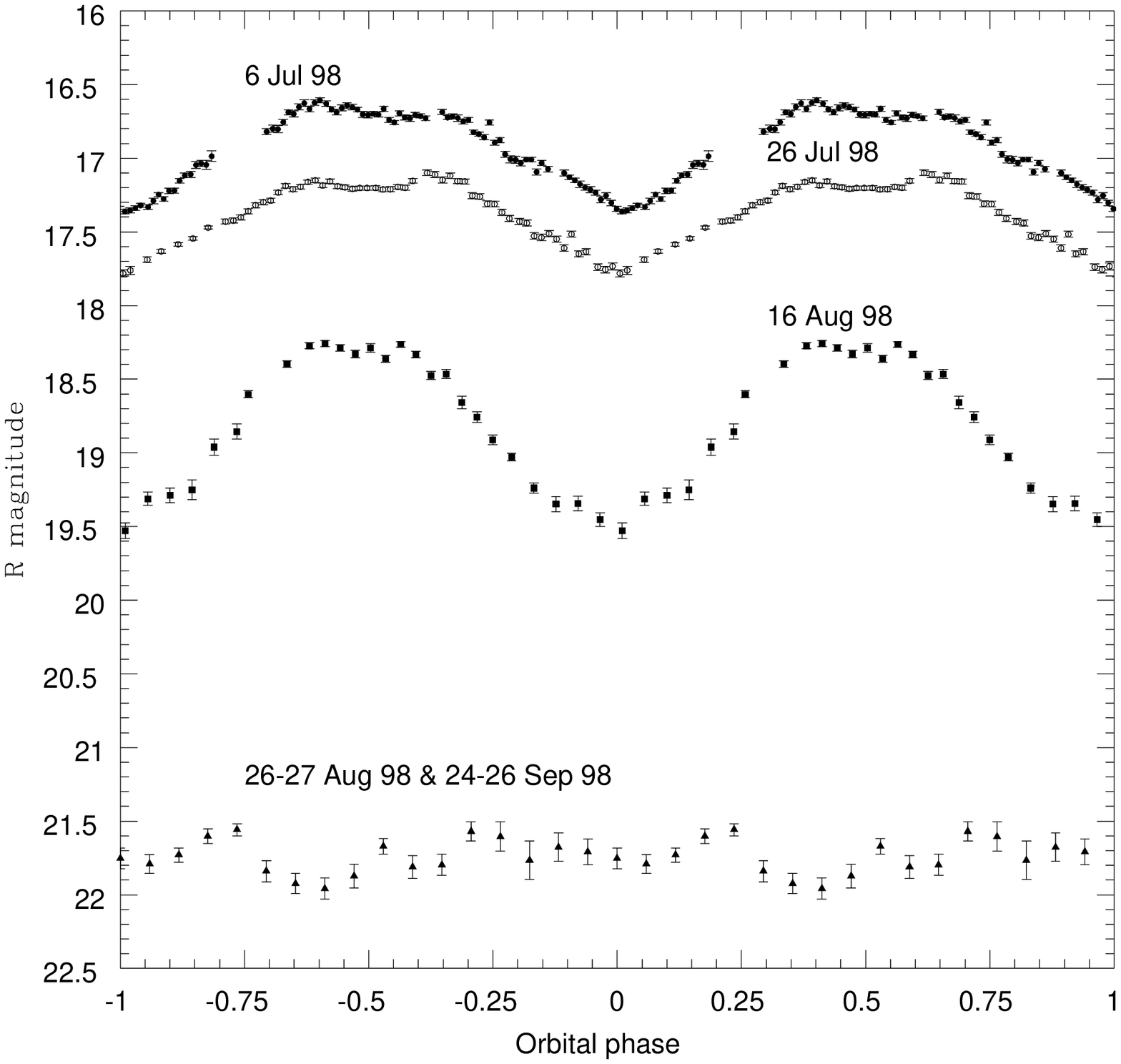}
\end{minipage}%
\begin{minipage}[c]{0.5\textwidth}
\centering
\includegraphics[width=5.5cm,angle=90]{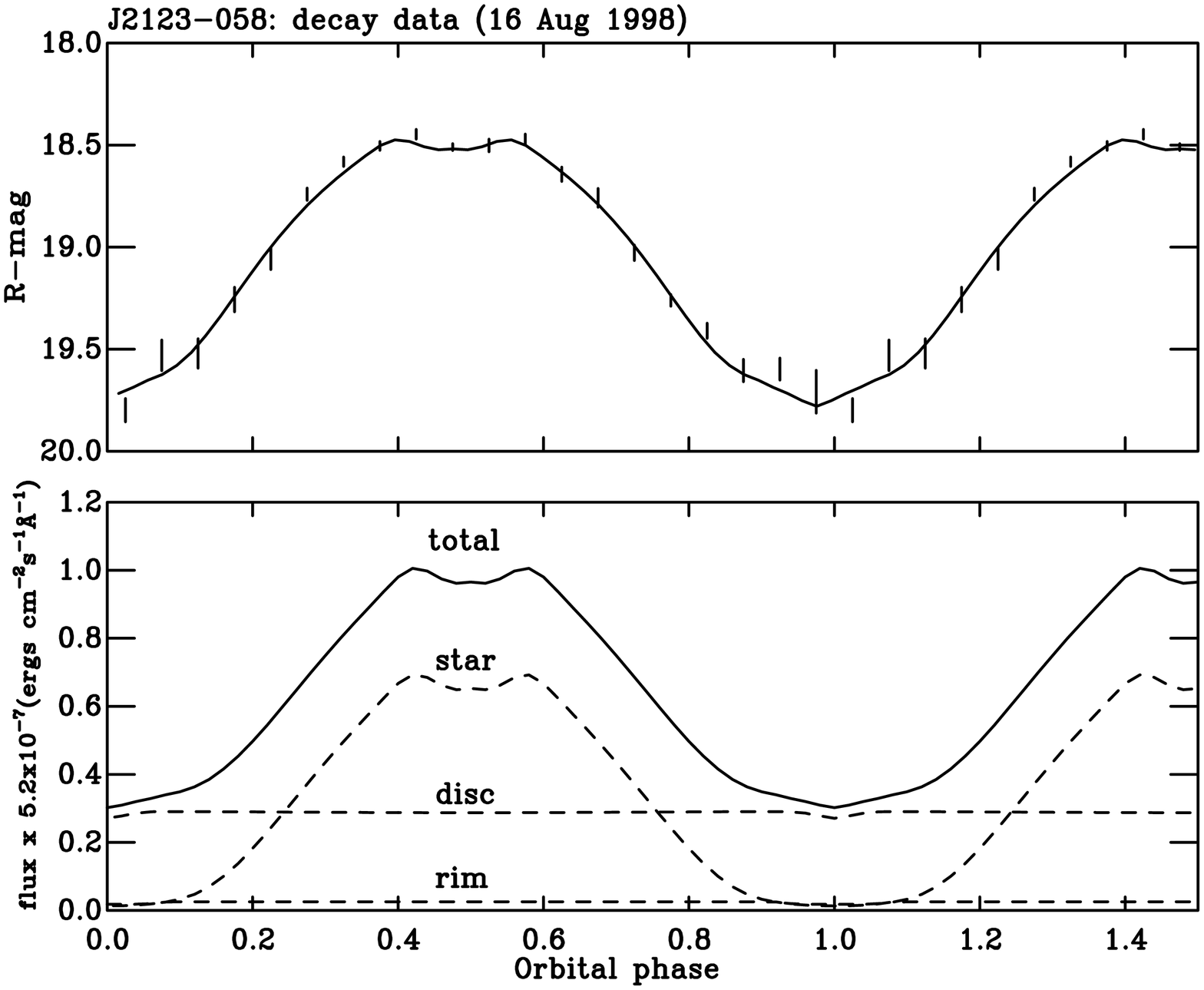} 
\end{minipage}
\caption{{\it Left:} Orbital light curves of J2123-058 (folded on $P=$6h) from close to outburst peak (July 1998) through to quiescence several months later, showing the dramatic lightcurve evolution as $L_X$ declines (Zurita et al 2000). {\it Right}: Model fit to the August decay data with the individual model components plotted below (Shahbaz et al 2003).
\label{2123lcs}}
\end{figure}

In addition to the outburst light curves of J2123-058, time-resolved
spectroscopy (Hynes et al 2001a) through two orbital cycles permitted
the application of doppler tomography in order to provide an {\it
image} of the accretion disc, but with surprising results.  Multiple
S-waves are easily visible in HeII line profiles, but the doppler maps
(fig~{\ref{2123maps}}) show this to be located in the lower left region
(on the {\it opposite} side of the NS from the
companion), and hence definitely {\it not} the donor's heated face.

\begin{figure}[t]
\begin{center}
{\bf a)} \includegraphics[angle=270,width=.337\textwidth]{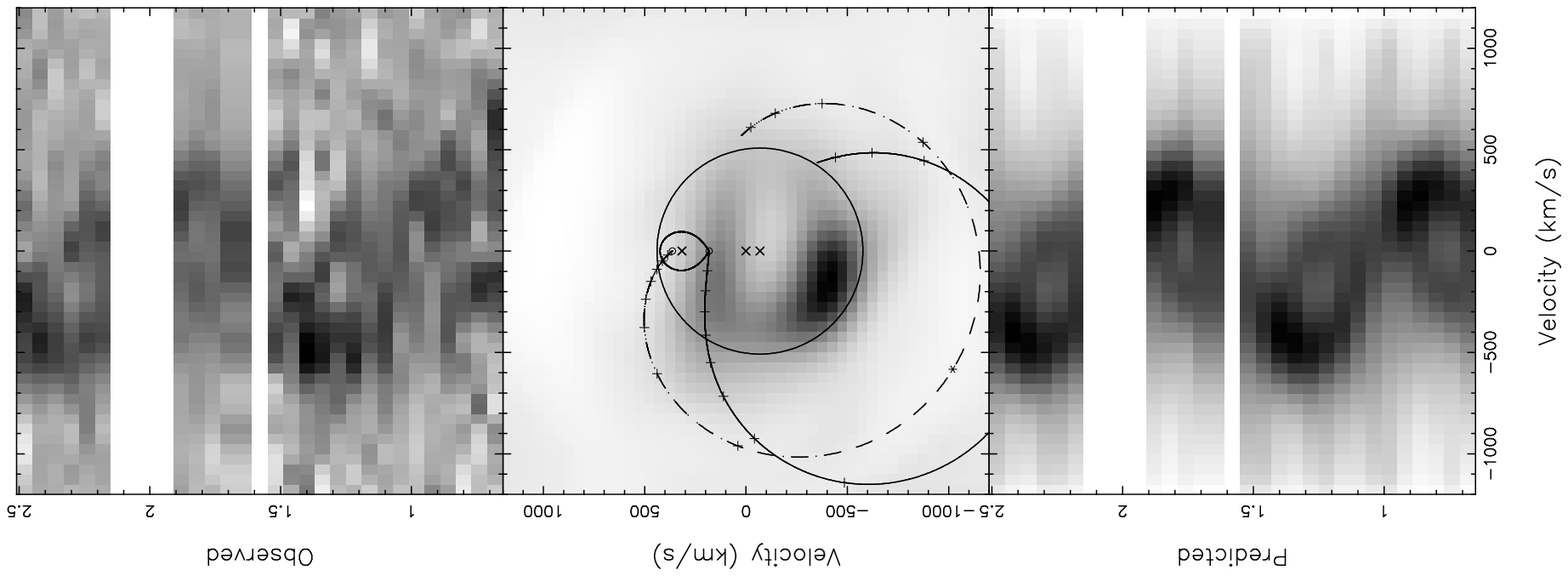}
\begin{minipage}[t]{.5\textwidth}
\vspace{-9mm}
\hspace*{0.5mm}
\begin{minipage}[b]{12pt}{\bf b)}\\\rule[0mm]{0mm}{57mm}\end{minipage}
\includegraphics[width=1.1\textwidth]{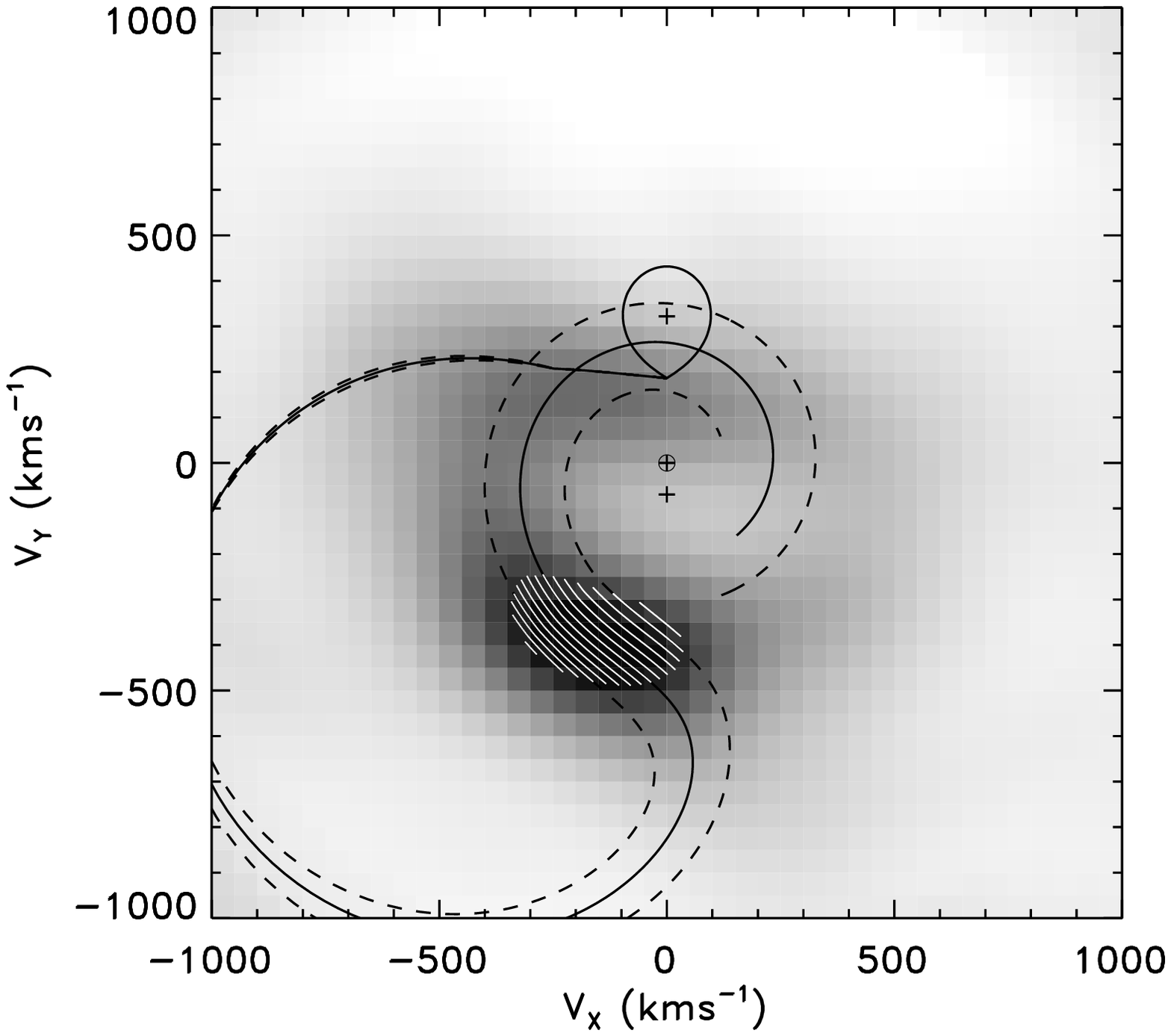}\\
\begin{minipage}[b]{12pt}{\bf c)}\\\rule[0mm]{0mm}{52mm}\end{minipage}
\includegraphics[width=\textwidth]{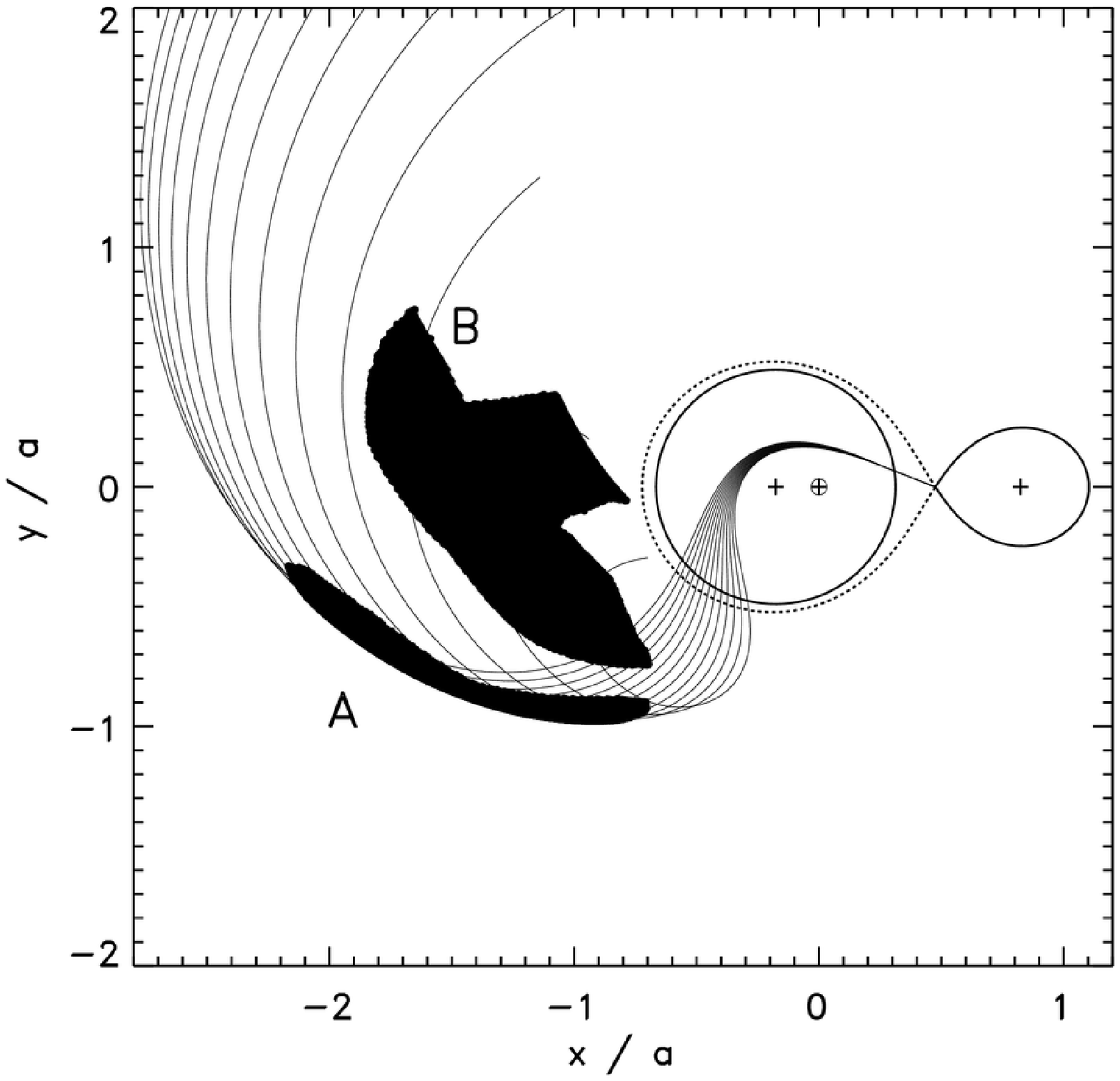}
\end{minipage}
\end{center}
\caption[]{Doppler tomography of J2123--058 during outburst (Hynes et al 2001a).  a) (from top) actual data, tomogram, reconstruction.  In
the centre panel, the solid line is the ballistic stream trajectory,
the dashed line the Keplerian velocity at the stream position and the
large circle is the Keplerian velocity at the disc edge.  b)
Comparison between magnetic propeller trajectories and the
tomogram. c) Spatial plot corresponding to (b).  Points in region A
produce emission with kinematics corresponding to the tomogram's white hatched
region, whereas those in B produce absorption
with the H$\alpha$ phasing and velocities.}
\label{2123maps}
\end{figure}

Combined with transient H$\alpha$ absorption between orbital phases
0.35-0.55, Hynes et al interpret J2123-058 as the NS analog of the SW
Sex phenomenon (a sub-class of CVs, see Hellier 2000).  While other
LMXBs, both transient (e.g. J0422+32, Casares et al 1995)
and steady (e.g. X1822-371, Harlaftis et al 1997; Casares
et al 2003) display their main emission at the
stream-disc impact region, there are now several LMXBs with similar
properties to J2123-058 (see fig~{\ref{CompFig}}).  While this region of the
doppler tomogram can be a result of stream overflow and
subsequent re-impact on the disc (Shafter et al 1988), it
requires a strongly flared disc in order to prevent the overflowing
stream producing absorption at {\it all} phases, and this is already
very tightly constrained by the light curve models described above.  A
more likely explanation is to invoke the {\it magnetic propellor}
model of AE Aqr (Eracleous \& Horne 1996; Horne
1999b) which involves a strong magnetic field anchored to
a rapidly spinning white dwarf which accelerates matter out of the
system and into a region beyond the compact object where the streams
collide (fig~{\ref{2123maps}}c).  However, J2123-058 involves a
rapidly spinning neutron star ($P_{spin}\sim$4ms; Tomsick et al
1999~\cite{tom99}) and would present far too small an area to be
effective, and so Hynes et al suggest that this process could still be
occurring but with a magnetic field anchored in the disc.

\begin{figure}[t]
\begin{center}
\includegraphics[width=.6\textwidth]{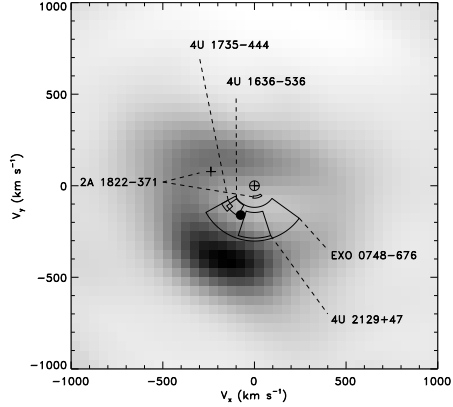}
\end{center}
\vspace{-1cm}
\caption[]{Comparison of J2123-058's tomogram with other systems. 
Boxes represent uncertainties
in velocity semi-amplitude and phasing.  The large point is the best
fit semi-amplitude and phase for J2123--058 (Hynes et al 2001b).}
\label{CompFig}
\end{figure}

\subsubsection{Millisecond pulsar transients}

Whilst the discovery of kHz QPOs in LMXBs (see chapter 2) finally
demonstrated that high $L_X$ LMXBs could indeed spin up their
neutron stars to very high rates (and hence are the ancestors of radio
millisecond pulsars, MSPs), it was not until 1998 that RXTE discovered
the first X-ray MSP, J1808.4-3658, with
$P_{spin}$=2.49ms (Wijnands \& van der Klis 1998).
X-ray bursts were also seen by SAX (in't Zand et al
1998) and this fast LMXB pulsar is in a 121min orbit
about a very low mass ($<$0.1\msun) secondary (Chakrabarty \& Morgan
1998).  This short-lived transient (only tens of days)
peaked at V$\sim$16.6 (Roche et al 1998), but was below
V$\sim$20 only 6 weeks later (Giles et al 1999).  However,
Giles et al did detect a ${\Delta}V\sim$0.1 modulation on $P_{orb}$
which was anti-phased with the pulsar and hence dominated by the
heated face of the cool donor.  Even in quiescence at V$\sim$21, Homer
et al (2001a) found a similar, sinusoidal modulation still
present, which was due to the heated face, and {\it not} the
double-humped ellipsoidal light curve seen in XRN,  i.e. the
secondary is {\it not} detectable.  This can be explained by the fact
that J1808.4-3658 has $F_X/F_{opt}\sim$10 in quiescence, compared to a
more typical $\sim$1000 in outburst i.e. the quiescent source is still
bright optically, which means that the Homer et al light curve is due
to a combination of the remnant (irradiated) disc and
the heated face of the donor.  

Furthermore Bildsten \& Chakrabarty (2001) have determined that the
donor has $M_2$=0.05\msun ~and $R_2$=0.13\rsun ~in order to fill its
Roche lobe, and hence it must be heated by the NS (see also Burderi et
al 2003), confirming that J1808.4-3658 is the progenitor of an MSP
binary.  The low average \mdot ~of $\sim$10$^{-11}$\msun$y^{-1}$
implies that such systems can be extremely long-lived and may still be
present as e.g. low $L_X$ sources in globular clusters (as seen in 47
Tuc, see chapter 8).  This class of object is extremely important as a
test-bed for understanding the magnetic properties of neutron stars
(Bhattacharya 2002), and will be enhanced with the extremely recent
discovery by RXTE of 4 more LMXB MSPs (Markwardt et al 2002; Galloway
et al 2002; Bildsten 2002; Markwardt et al 2003; Kirsch \& Kendziorra
2003), all with $P_{spin}\sim$2-5ms.

\subsection{Infrared spectroscopy of LMXBs}

It is an unavoidable consequence of the galactic distribution of LMXBs
(White \& van Paradijs 1996) in which half of all LMXBs are within 20
degrees of the Galactic Centre, that the majority of these are heavily
obscured in the optical and hence are accessible only in the IR (since
the {\it V}/{\it K} extinction ratio is $\sim$10).  The most luminous
of these form a sub-group known as the ``Galactic Bulge Sources'' (or
GBS) which lie within a strip of {\it l}=$\pm$15$^\circ$, {\it
b}=$\pm$2$^\circ$ about the Galactic Centre.  Several of the GBS
exhibit QPOs, but rarely X-ray bursts (see chapter 3), and no clearly
defined $P_{orb}$, hence their nature as LMXBs or HMXBs was unclear in
many cases.  Consequently, the accurate X-ray locations for the GBS
were surveyed by Naylor, Charles \& Longmore (1991) with the first
generation of IR arrays.  While the technology has improved over the
last decade, IR spectroscopy is still limited to much brighter sources
than in the optical, and so much remains to be done.  Here we discuss
the advances made on particular sources (e.g. Bandyopadhyay et al
1997, 1999), key properties of which are in table~{\ref{tab:IRLMXBs}
together with details of other highly obscured LMXBs (apart from Sco
X-1 and Cyg X-2 which are included for reference)}:

\begin{itemize}

\item GX1+4 (X1728-247) is unusual in being in 
a region of low reddening, where the X-ray source is coincident with a
bright ({\it K}=8.1) M6III star at $d\sim$3--6~kpc and hence
$L_X\sim$10$^{37}$\ergsec (Chakrabarty \& Roche 1997).
$A_V$ is steady (at 5.0$\pm$0.6), but the X-ray
column is variable, implying the presence of a wind.  Its H$\alpha$
emission is not uncommon for such an object, and the identification
was only secure when the H$\alpha$ flux was found to be pulsing with
the same period (114s) as in X-rays (Jablonski et al
1997).  While GX1+4 was, for 25 years, the brightest hard
X-ray source in the Galactic Centre region (and the pulsar was
spinning up), it turned ``off'' in the 1980s and is now on at a low
level (and spinning down).  Most HMXBs display an orbital
modulation superposed on the pulsar's long-term ${\dot P}$ trend,
but not GX1+4.  There have been suggestions of a $\sim$304d variation
in the accretion torque history (Cutler et al 1986) and
spin-down rate (Pereira et al 1999), but there is no
correlation with the X-ray flux and so these claims require
confirmation.  Assuming a NS mass of 1.4\msun and requiring
the M6III to fill its Roche lobe implies that $P\geq$100d.

GX1+4 is of interest as the {\it only} NS symbiotic system
(but note 4U1700+24, Masetti et al 2002, which may be similar,
although much less luminous), and hence a potential progenitor of very
wide ($P\gg$10d) radio pulsar binaries.  These must have been very
wide LMXBs and hence required K or M giant donors.  Such rare systems
(with a high {$\dot M$}) would appear just like a symbiotic.

\item GX13+1 was identified with a $K\sim$12 star by Naylor et al
(1991) on the basis of its variability and strong Br$\gamma$ emission
(fig~{\ref{gx13ir}}).  More importantly the spectrum also shows CO
absorption bands typical of cool (K--M) stars, and the
presence of $^{13}$CO indicates that it is evolved.  At
$d\sim$9kpc and E(B-V)=5.7, then M$_K$=--4.6 and so it must be
$\sim$K2--M5III.  Possible periodicities in the range 13--20d have
been suggested (Garcia et al 1992; Bandyopadhyay et al
2002) which would be consistent with such a
classification, but further advances in understanding GX13+1 require
an IR radial velocity study, which is now feasible, although
observationally challenging.

\begin{figure}
\includegraphics[width=6cm,angle=90]{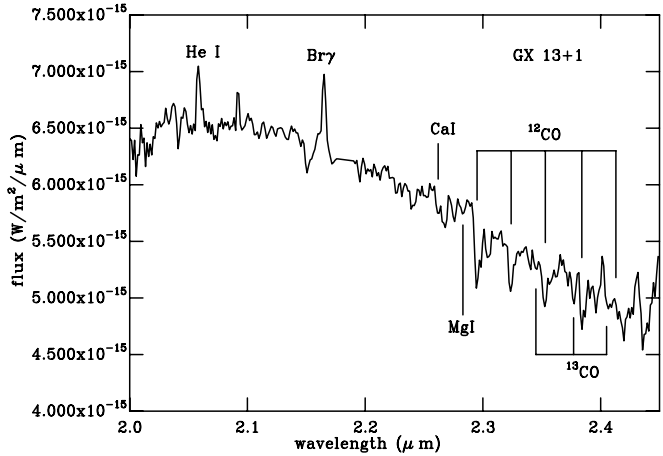}
\caption{
K-band spectrum of GX13+1 (Bandyopadhyay et al 1999)
showing strong HeI and Br$\gamma$ emission, together with CO
absorption bands.  Note the presence of $^{13}$CO which indicates it
is both cool and evolved.}\label{gx13ir}
\end{figure}

\item GX17+2 is a bright Z source which also bursts (and is hence a
NS).  It benefits from an accurate radio location (Hjellming
1978) which aligns with NP Ser, a V$\sim$17, completely
normal G star.  This conundrum was settled by an HST/NICMOS image
which revealed that NP Ser is actually outside the radio error circle
(by 0.7 arcsecs), while two much fainter (H$\sim$20, 21) stars are
consistent with the radio location (Deutsch et al 1999).
Keck observations (Callanan et al 2002), however, reveal a
K=15 star at that position, $\sim$3.5-4 mags brighter than the NICMOS
stars!  Furthermore, K-band monitoring by Bandyopadhyay et al
(2002) shows that this object displays large
amplitude variability once corrections are made for (the unresolved)
NP Ser.  What is remarkable about GX17+2 is that (according to
RXTE/ASM monitoring) the X-ray flux level is extremely high {\it and}
steady, in spite of the huge IR fluctuations.

\item the two SXTs, X1608-522 and X1630-47, are unusual in that they both 
have semi-regular outbursts on timescales of $\sim$600 days
(Augusteijn et al 2001; Wachter et al 2002;
and references therein).  They both suffer from extinction, so
their optical counterparts are faint, even in outburst, but both have
now been observed in the IR, with X1608-522 recently revealing a 0.54d
(single peak, i.e. X-ray heated) modulation during an extended low
X-ray intensity state following an outburst.  Such states are compared
with Z Cam ``standstills'' which are seen in CVs and GX339-4 (Kong et
al 2002), and have also been seen in Aql X-1, another frequently
outbursting, neutron star SXT.

\end{itemize}
{\small
\begin{table*}
\begin{center}
\caption{\label{tab:IRLMXBs}LMXB IR Properties}

\begin{tabular}{lccclll} 
\hline
{\em Source}  & {\em b$^{II}$} & $K$ & $P$ & \multicolumn{2}{c}{IR Spectrum} & Notes \\
  & & & (d) & {\em emission} & {\em absorption} & \\

\hline

Cir X-1$^{1,2}$ & 0 & 9--12 & 16.54 & Br$\gamma$, HeI, II & - & Super $L_{Edd}$, highly eccentric orbit \\
X1630-47$^{3}$ & +0.3 & 16.1 & - & - & - & $\sim$600-690$^d$ recurrent BH transient \\
GX1+4$^{4,5}$ & +4.8 & 8 & $\sim$300 & Br$\gamma$ & M6III & symbiotic, X-ray pulsar (114s) \\
\\
{\it Z-sources:} & & & & \\ 
Sco X-1$^{4,6}$ & +24 & 11.9 & 0.79 & Br$\gamma$, HeI, II & - & ``steady'' prototype LMXB\\
GX340+0$^{7}$ & -0.1 & 17.3 & - & - & - & luminous bulge source \\
Sco X-2$^{6}$ & +2.7 & 14.6 & 0.93/14.9 & Br$\gamma$, HeI & - & `` \\
GX5-1$^{8,9}$ & -1.0 & 13.5 & - & Br$\gamma$, HeI & - & `` \\
GX17+2$^{10-12}$ & +1.3 & 14.9--18.5 & $\sim$20 & - & - & {\bf not} NP Ser \\
Cyg X-2$^{7}$ & -11.3 & 13.3 & 9.8 & - & A9III & IMXB \\
\\
{\it atolls:} & & & & \\
X1608-522$^{7,13}$ & -0.9 & 16.5 & 0.54 & - & - & NS transient  \\
GX9+9$^{7}$ & +9.0 & 16.0 & 0.17 & - & - & v short period \\
GX3+1$^{14}$ & +0.8 & 15.1 & - & - & - & candidate ID \\
GX9+1$^{14}$ & +1.1 & 16.2 & - & - & - & candidate ID \\
GX13+1$^{4,6,12,15}$ & +0.1 & 11.9-12.3 & $\sim$25d & Br$\gamma$, HeI & CaI, MgI, & (see fig~{\ref{gx13ir}}) \\
 & & & & &  $^{12}$CO, $^{13}$CO \\
\hline
\end{tabular}
{\footnotesize
$^1$Johnston et al 1999; 
$^2$Clark et al 2003;  
$^3$Augusteijn et al 2001;
$^4$Bandyopadhyay et al 1997;
$^5$Chakrabarty \& Roche 1997;
$^6$Bandyopadhyay et al 1999;
$^7$Wachter 1998;
$^8$Jonker et al 2000;
$^9$Bandyopadhyay et al 2003;
$^{10}$Deutsch et al 1999;
$^{11}$Callanan et al 2002;
$^{12}$Bandyopadhyay et al 2002;
$^{13}$Wachter et al 2002;
$^{14}$Naylor et al 1991;
$^{15}$Charles \& Naylor 1992
}
\end{center}
\end{table*}
}

\subsection{Long Periods and Disc Structure in High Inclination LMXBs \label{sect:lps}}

The first X-ray dipper discovered, X1916-053 (Walter et al 1982),
produced one of the first orbital periods of any LMXB.  With type I
X-ray bursts indicating the presence of a neutron star, and the
ultra-short $P_{orb}$ of 50 mins implying that the mass donor was a
$\sim$0.1M$_\odot$ degenerate dwarf, then $q\sim$ 15 and hence it
should also be susceptible to the SU UMa-type disc precession
(section~{\ref{sect:shumps}}).  Indeed an X-ray quasi-periodicity of
$\sim$199d had already been suggested (Smale \& Lochner 1992).

In spite of its faintness (B$\sim$21; due to a combination of
reddening and the low-mass donor) and crowding, an optical analogue of
the X-ray dip period was found, but remarkably it was 1\% longer
(3027.6s versus 3000.6s)!  Both periods are now precisely known and
the difference is highly significant (Callanan et al 1995; Chou et al
2001; Homer et al 2001b), but debate continues as to which is orbital.
Classical SU UMa-type behaviour would imply it is the X-ray, and
precession induces the slightly longer optical modulation.  However,
unlike in SU UMa systems (where the superhump period actually drifts
during the $\sim$2-week superoutburst), the optical period in
X1916-053 was extremely stable, suggesting that it might be the
orbital period.  Other models have also been invoked (e.g. a triple
system, Grindlay et al 1988, Grindlay 1989).

Furthermore, detailed X-ray and optical light-curves
(fig~{\ref{1916Xlc}}; see Homer et al 2001b; Chou et al 2001) reveal
that the morphology evolves dramatically over several days, on a
timescale similar (but not necessarily equal) to the X-ray-optical
beat period (fig~{\ref{1916Xlc}).  The X-ray dipping is known to be
due to azimuthal structure in the disc edge (e.g. around the
stream-disc impact region) viewed at an appropriately high $i$ (Parmar
\& White 1988).  Any variation in this structure, or in the angle it
presents relative to our line of sight, would cause the observed light
curve to change.  This has led to speculation that the disc in
X1916-053 not only precesses but is warped, an effect expected due to
X-ray irradiation (Wijers \& Pringle 1999; Ogilvie \& Dubus 2001).
This is presumably similar to the 35d on-off cycle in Her X-1, where
the tilted disc precesses relative to the observer (see e.g. Schandl
\& Meyer 1994).  This cycle is not precise, it has some ``jitter'', as
is also seen in the precession of SS433 (Margon 1984).

\begin{figure}
\includegraphics[angle=-90,scale=0.15]{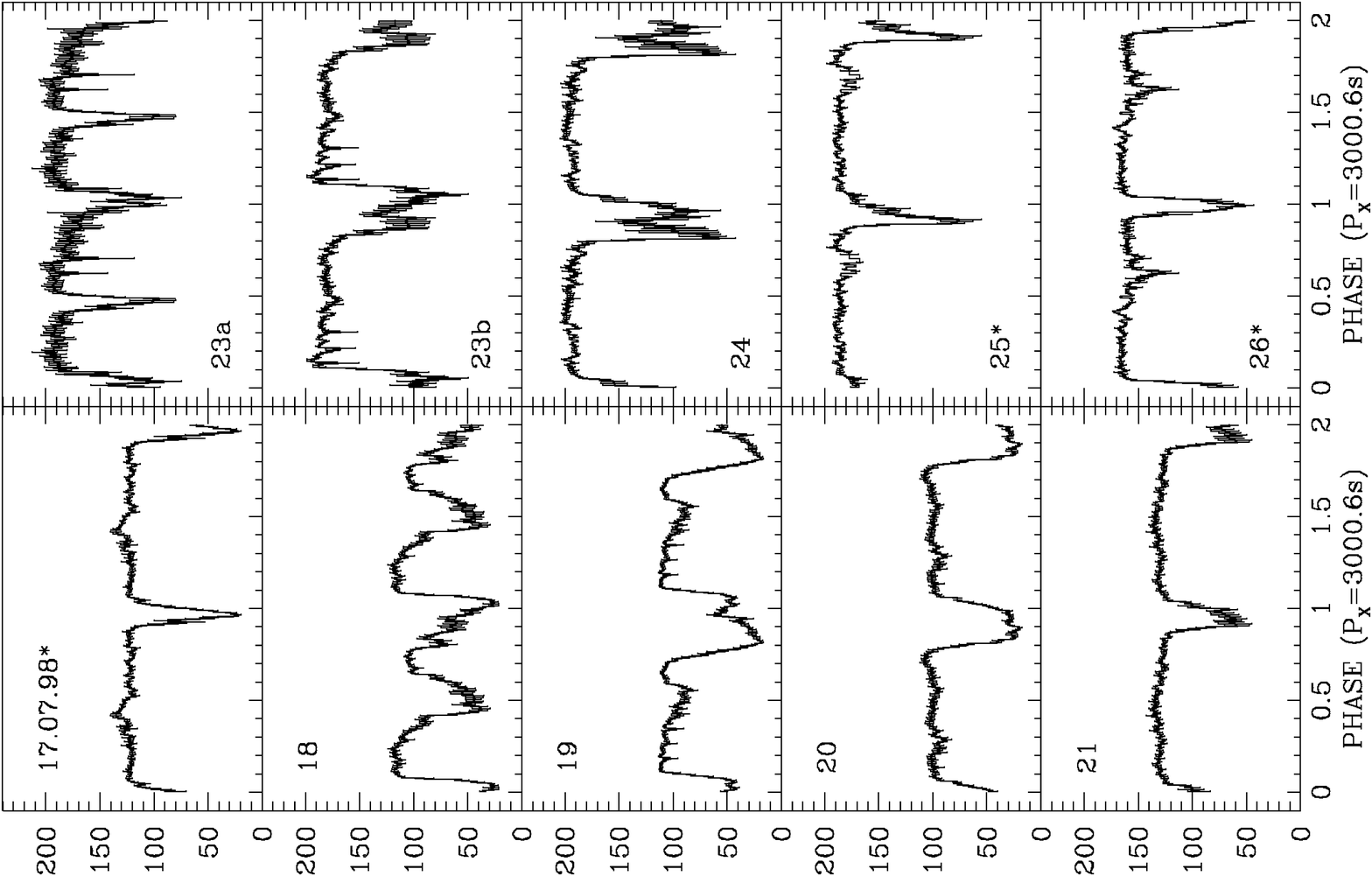}
\caption{RXTE PCA X-ray light curves of X1916-053 folded on the X-ray period of 3000.6s, in which the dipping structure clearly evolves on a timescale of days, eventually repeating (adapted from Homer et al 2001b). \label{1916Xlc}}
\end{figure}

Such a light curve modulation can also be explained by invoking
a triple system (producing dips predominantly when the mass transfer
rate is enhanced), which would have to be formed in a globular
cluster core which has subsequently evaporated (Grindlay et al 1988).
A further problem is the stability of the triple given the observed
periodicities, and this model is now deemed unlikely.  The effects
observed are far closer to those expected from the precessing
disc model related to the well-established high $q$.  Initially
the stability of the optical period was considered a difficulty, but
since the mass transfer in this LMXB is basically steady (unlike in
SU UMa systems) it is possible that the instability
will also stay fixed at a single period, i.e. X1916-053 is a {\it
permanent superhumper}.  An alternative is to consider X1916-053 as a
{\it negative superhumper} in which the optical period is indeed
orbital, and the X-rays are now the superhump.  While this is seen
in a number of CVs (see e.g. Patterson 1999) the effect is not
well understood (Retter et al 2002).

With its short $P_{orb}$ and accessible ($\sim$days) light curve
evolution, X1916-053 is an ideal system in which to study accretion
disc edge structure and how it might be influenced by
irradiation-driven warping.  Such results are likely to have
application well beyond this immediate field.  Possibly similar
behaviour has been reported in the BHC X1957+11 (Hakala et al
1999) and the system MS1603+2600 (Hakala et al
1998), the latter at high latitude and the nature of the
compact object is still controversial (see e.g. Mukai et al
2001).

\subsubsection{The halo transient, XTE~J1118+480}

This topic has been taken even further with extensive observations of
the SXT J1118+480 (Remillard et al 2000).  With its low
extinction and relative proximity ($\sim$1kpc) the late decline phase
was monitored in unprecedented detail (Zurita et al 2002a).  As
expected once the disc contribution faded, the ellipsoidal
modulation became visible (figure~{\ref{j1118sh}}).  However, the
accretion disc component was still very significant ($\sim$50--70\%),
and temporal analysis revealed that, in addition to the ellipsoidal
variation (which appears to peak at $P_{orb}/2$), there is a further
modulation at a slightly longer period (visible as a distorting
``wave'' in the nightly light curves).  Subtracting a theoretical
ellipsoidal fit from the data leaves this unusually structured
modulation, an effect that continues well past the original outburst
(as also now seen in CVs, e.g. Patterson 1995).  More surprisingly
here, the period excess is extremely small, at only $\Delta\sim$1\%.

\begin{figure}
\centering
\begin{minipage}[c]{0.5\textwidth}
\centering
\includegraphics[width=5cm]{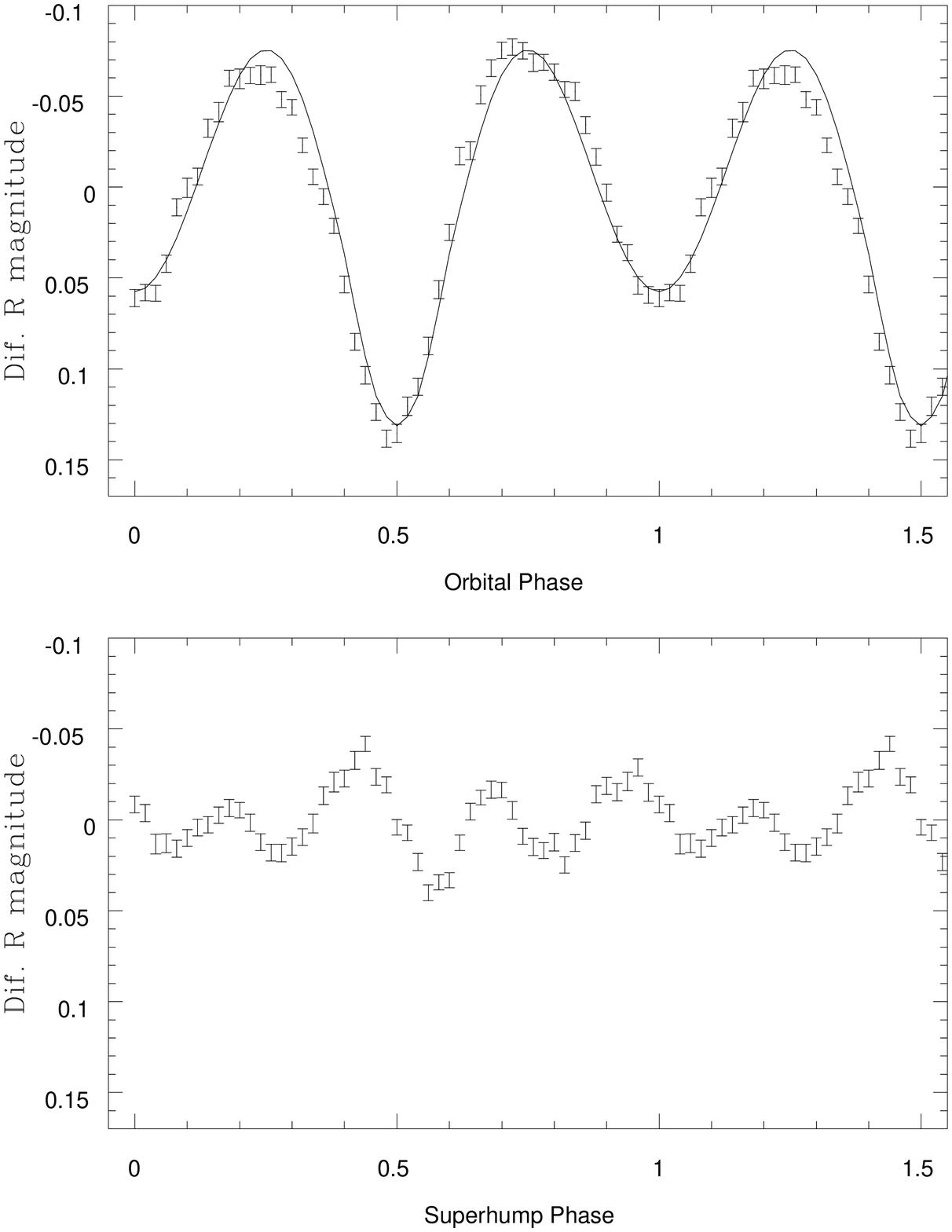}
\end{minipage}%
\begin{minipage}[c]{0.5\textwidth}
\centering
\includegraphics[width=6cm,angle=0]{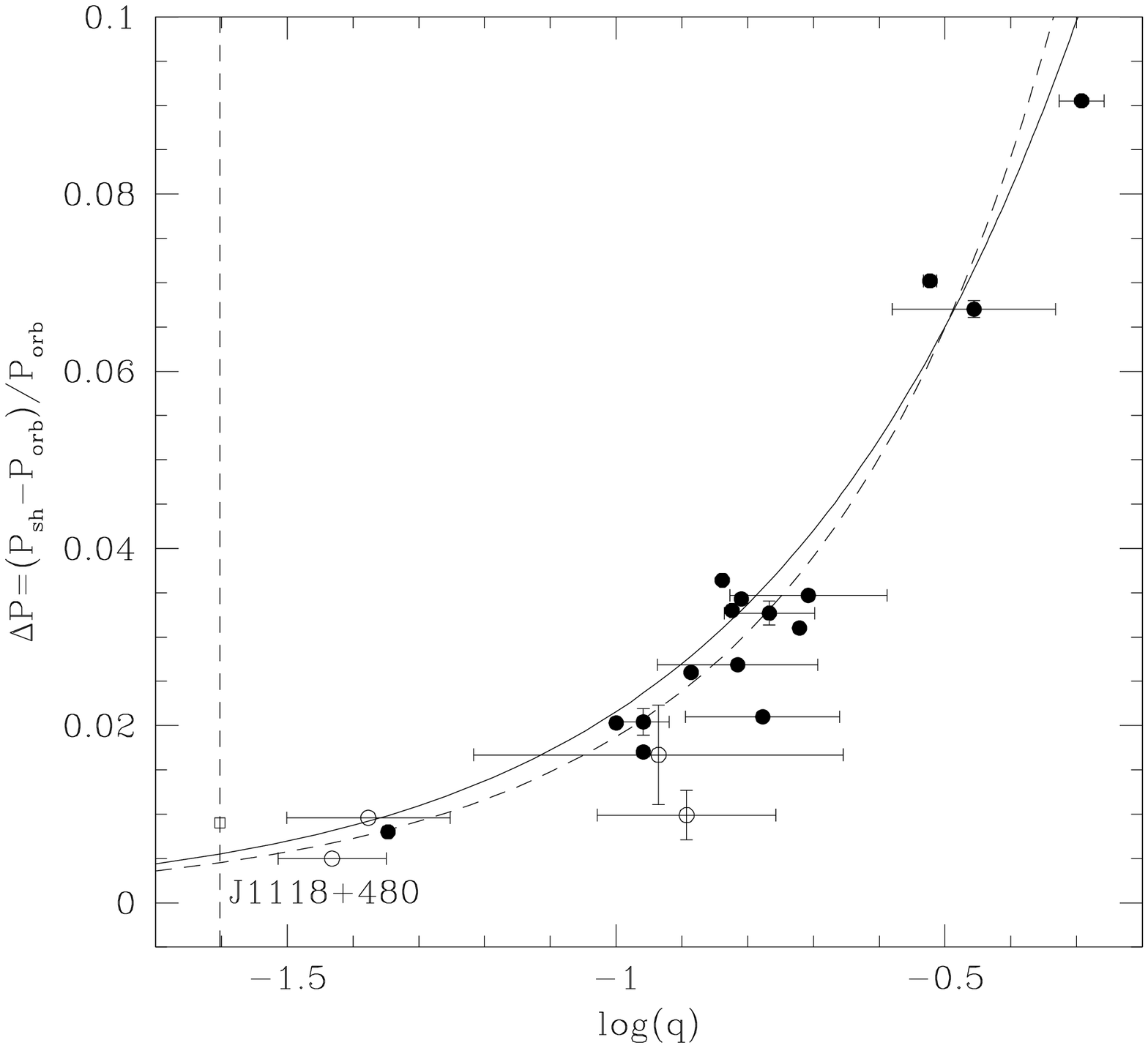} 
\end{minipage}
\caption{{\it Left}: 
Detrended lightcurve (upper) of J1118+480 folded on
$P_{orb}$=4.078h, together with an ($i$=75$^\circ$ and $q$=25) ellipsoidal model, with residuals plotted below after
folding on $P_{sup}$=4.092 hours.  
{\it Right}: Combined results on LMXBs (J1118+480, X1916-053) and CVs (Patterson 2001) that display superhumps, where $\Delta$ is plotted against $q$.  Models are plotted as solid (Patterson) and dashed (Osaki 1985~\cite{osa85}) curves, but should not be valid below $q=0.025$ (Zurita
et al 2002a). \label{j1118sh}}
\end{figure}

However, such a small $\Delta$ implies (equation (5.6)) that $q$ must
be extremely high ($\sim$40), and hence this interpretation must be
treated with great caution (fig~{\ref{j1118sh}}).  At such
extreme $q$ values the classical 3:1 resonance would be replaced by
the 2:1 resonance (Whitehurst \& King 1991).

\subsection{Special Cases}

In this final section we discuss several high profile sources whose
classification as either HMXB, LMXB or even binary is still a subject of
controversy, mostly due to the absence of any direct signature of the
mass donor.  However, in 3 of these cases there is now
circumstantial evidence that the secondaries are quite massive,
but the very high implied \mdot gives them properties akin
to LMXBs, so they transcend a simple classification.

\subsubsection{SS433: the link with ULXs?}

SS433 is the original source for studying the detailed physical
processes involving relativistic jets on a galactic scale, and is
still the only continuously emitting galactic micro-quasar (see
chapter 9).  The key feature of SS433 is its 160d precessing jet,
observed for several decades (in optical, IR and X-ray) as periodic
emission-line shifts which can be fitted to high precision by the {\it
Kinematic Model} (see Margon 1984 for details and a full historical
review).  However, in spite of more than 20 years of detailed study,
remarkably little is known about the fundamental system properties of
SS433.  This is a consequence of two factors: (i) the enormously
powerful and broad, stationary and moving emission lines (which first
brought it to attention) obliterate any spectral features of the
companion; (ii) the high extinction which has severely biassed most
optical spectroscopy towards the red (which Goranskii et al 1998 show
is in any case dominated by an erratically varying component) and
hence away from where intrinsic spectral features of a massive donor
might be observed.  The properties of the microquasars, such as SS433
and GRS1915+105, are of great importance in furthering our
understanding and interpretation of the ultra-luminous (ULX) sources
in nearby galaxies (see chapters 9, 12 and 13).  While SS433 appears
to us as an extremely low $L_X$ ($\sim$10$^{36}$erg~s$^{-1}$) source
compared to GRS1915+105 (which peaks at $\sim$10$^{39}$erg~s$^{-1}$,
and possibly up to a hundred times greater than this, once appropriate
correction for extinction has been made, e.g. Greiner et al 1998), it
is possible that the high $i$ and extremely high \mdot of SS433 are
preventing us from viewing its intrinsic $L_X$.  In any case the
observed $L_X$ is far exceeded by the jet power, as is true in pulsar
wind nebulae, but not in the other microquasars (Safi-Harb 2003).

Nevertheless, the orbital period is known (13d) and it is eclipsing in
optical and X-ray light curves, but with no signature of the mass
donor there exists no constraint on the dynamical masses comparable to
those of the XRN described earlier.  Indeed, there are published
``determinations'' of the SS433 compact object mass ranging from
0.8\msun~(d'Odorico et al 1991) to 62\msun~(Antokhina \& Cherepashchuk
1985)!  These are all based on radial velocity curves of the HeII
emission line, of which the most detailed study is that of Fabrika \&
Bychkova (1990), who used $\sim$4 yrs of 6m data taken around
$\Psi\sim$0 (which gives the lowest scatter in their radial velocity
curve), when the disc normal was closest to our line of sight.
Assuming $e$=0 they obtain $f(M)\simeq$8\msun, which implies a
10\msun~donor for a NS compact object, or 15\msun~for a 6\msun~BH (and
which would also better explain the extremely high
\mdot$\sim$10$^{-4}$\msun$y^{-1}$, see also Fuchs et al 2003).

But perhaps the best attempt so far to directly observe the mass donor
has come from Gies et al (2002) who, using the
interpretation that SS433 is a binary embedded in an expanding thick
disc wind that explains the equatorial radio disc of Blundell et al
(2001) and the stationary lines, also decided to observe
SS433 spectroscopically at precessional phase, $\Psi\sim$0, and in the
blue where the expected high mass donor would be more easily
detectable (and has historically been avoided due to the consequent
need to use larger telescopes).  During eclipse, Gies et al report the
possible detection of (weak) spectral features that are typical of an
A supergiant, which is consistent with photometry (Antokhina \&
Cherekpashchuk 1987; Goranskii et al 1997)
that shows the donor to be cooler than the disc, which still dominates
even in the blue.  The velocities of these features and known
orbital phase are combined with an assumed $e$=0 and Fabrika \&
Bychkova's K-velocity for HeII (assumed to represent the disc) to
yield $q=M_2/M_X$=0.57, $M_X$=11$\pm$5\msun, $M_2$=19$\pm$7\msun,
$R_2$=31$\pm$3\rsun (assuming it is Roche-lobe filling), hence
requiring a BH of similar mass to those found in the XRN.  However,
this result must be treated with caution until a full radial velocity
curve has been obtained, but this will be extremely difficult given
the large contribution from the bright and hot disc.

There also now exists a large database of SS433 V-band photometry
(from 1979-1996) that Fabrika \& Irsmambetova (2002) have reanalysed
by using (mostly) radio coverage to separate into active and passive
states, which were then examined on both orbital and precessional
phases.  The most remarkable result is that the largest
(optical/radio) flares occur almost entirely at orbital $\phi\sim$0.3
(with a few at $\sim$0.8), which implies a non-circular orbit!
However, the well-known nodding motions in the jet (Katz et al
1982~\cite{kgm82}) set an independent upper limit of $e<$0.05, but
Fabrika \& Irsmambetova point out that an eccentricity as small as
$e\simeq$0.01 would be enough to modulate the Roche volume and thereby
perturb the disc; it being the disc, and not \mdot~ which is already
extremely high, that will modulate the flux, following an $\sim$1d
delay in propogating the \mdot~ change through the disc.  Such a small
value of $e$ would be undetectable in current optical spectroscopic
studies.

\subsubsection{Cyg X-3: a Wolf-Rayet-BH X-ray binary}

Cygnus X-3 is a bright HMXB with the exceptionally short $P_{orb}$ of
4.8h. It is also a strong radio source showing extensive outbursts
associated with mass ejection events in a relativistic jet (Geldzahler
et al, 1983). Because of its high extinction, it took the advent of IR
spectroscopy to identify the nature of the mass donor when van
Kerkwijk et al (1992) showed the IR emission lines to be arising from
the strong stellar wind of a Wolf-Rayet (WN7) star, the only one
currently identified with an X-ray binary. Subsequently, van Kerkwijk
(1993), van Kerkwijk et al (1996) and Fender et al (1999) carried out
detailed studies of the IR line profiles and explained the double and
P-Cygni shapes to be associated with the high \mdot~ from the WR star
and {\it not} the jets.  Mid-IR ISOPHOT 2--12$\mu$ observations
(Koch-Miramond et al 2002) infer an expanding wind with
\mdot$\sim$1.2$\pm$0.5$\times$10$^{-4}$ \msun y$^{-1}$ from the
$\sim$WN8 donor.

X-ray spectroscopy by Paerels et al (2000)
using the {\it Chandra} HETGS
revealed a rich spectrum of photoionization-driven excited lines. An
overall net redshift of $\sim$750\kms, independent of binary phase,
was interpreted as an ionized region detached from the main mass
outflow. Furthermore, detailed {\it Chandra} imaging (Heindl et al,
2003) revealed an X-ray structure extending some 16 arcsecs
from the nucleus, which could be evidence
for the strong stellar wind impacting the ISM, thereby producing
bremsstrahlung X-ray emission.

The nature of the compact object within Cyg X-3 remains a puzzle. No
direct pulsations have ever been detected which would indicate a
NS. The presence of radio relativistic jets encourages many to include
it in the galactic microquasar group, and hence, by implication,
suggest that the compact object is a BH.  And while van Kerkwijk
(1993) and van Kerkwijk et al (1996) interpreted the IR line profile
variations as due to geometrical variations in the X-ray illuminated
hemisphere of the WR wind (supported by the correlation of line
strength and wavelength shift), Schmutz et al (1996) obtained higher
resolution spectra where the profiles did not follow the variation
expected of this model.  Instead, they interpret them as due to the
straightforward orbital motion of the WR star which then implies
$f(M)$=2.3\msun.  With WN7 star masses in the range 10--50\msun
(Cherepashchuk \& Moffat 1994), or more conservatively 5--20\msun
(Massey 1982), and $i$ in the range 30--90$^\circ$ then $M_X>$7\msun,
which makes Cyg X-3 an important system as potentially the {\it only}
currently identified WR/BH system (although two further candidates
have been proposed in the LMC, Wang 1995, and in IC10, Clark \&
Crowther 2003), suggesting either a rare formation mechanism (e.g. van
den Heuvel \& de Loore 1973) or a short lifetime, which might be
expected given the extremely high inferred \mdot.  A more detailed
analysis of the Fender et al data is given in Hanson et al (2000)
which reveals, during outburst, an absorption component in HeI which
is out of phase (by $\sim$0.25 cycle) with the emission lines.  They
present a compelling case, for interpreting this as the radial
velocity curve of the WR wind, yielding similar masses to those above.

\subsubsection{Cir X-1: HMXB or LMXB?}

Since its discovery in the early 1970's (Margon et al. 1971) the
nature of the X-ray binary Cir X-1 has remained elusive, in spite of
the known $P_{orb}$=16.6d inferred from the modulation of the X-ray
(Kaluzienski et al. 1976), near-IR (Glass 1994) and radio fluxes
(Haynes et al. 1978) and the discovery of Type I X-ray bursts
identifying the compact object as a NS (Tennant et
al. 1986). High extinction has made spectral classification of
the donor impossible: the best optical spectra to date are
dominated by asymmetric, HI, HeI emission lines (Johnston et
al. 1999; Johnston et al. 2001), with no photospheric features
evident.

The lack of spectral type has made determination of the accretion mode
difficult. Murdin et al. (1980) interpreted the X-ray and radio
modulations as evidence for direct accretion from a massive stellar
wind. Indeed the X-ray behaviour of Cir X-1 is similar to that of the
16.65d BeX A0538-66, in which wind-fed accretion is enhanced by
pseudo Roche-lobe overflow at periastron passage (Charles et
al. 1983).

Conversely, Johnston et al. (1999) suggest that a similar scenario is
also possible with an intermediate mass (3-5\msun)
companion. Furthermore, spectral-timing studies of Cir X-1 with
RXTE/PCA (Shirey et al. 1999) show Z-source X-ray QPO behaviour,
suggesting an LMXB classification (cf. van der Klis 1995). However,
the proposed association with the young SNR G321.9-0.3, which would
support a low-mass classification has recently been disproved by a
proper motion study (Mignani et al. 2002). Detailed IR spectroscopy
(Clark et al. 2003) reveals a spectrum that superficially resembles
that of a high-luminosity mid-B supergiant, however it is concluded
that the true spectrum of the donor remains obscured by the emission
from the disc or accretion driven outflows.

Clearly a consensus on the nature of Cir X-1 has yet to be reached.

\subsubsection{CI Cam: a fast transient B[e] HMXB?}

In 1998 RXTE detected an extremely bright ($\sim$2 Crab) new
transient, J0421+560 (Smith et al, 1998) with a positional error
circle that included the known, bright (V$\sim$11) B[e] supergiant CI
Cam. Optical spectroscopy at the time of outburst (Wagner \&
Starrfield 1998) revealed HeII emission superposed on the already rich
emission line spectrum, which strongly suggested CI Cam was
responsible for the X-ray source. Contemporaneous photometry (Hynes et
al, 2002c) revealed CI Cam to be 2-3 magnitudes brighter than usual,
and this combined with a radio detection by Hjellming \& Mioduszewski
(1998) confirmed the identification.  However, the extremely rapid
decay ($\sim$2d) of the outburst was highly unusual, with only V4641
Sgr of the XRN described earlier exhibiting similarly fast variations.

Hynes et al (2002c) have presented a major review of this system, but
there are still some major questions left unanswered.  It has been
demonstrated that CI Cam is a sgB[e] which underwent a major mass
outburst phase, and that this ejected material interacted with a
compact object (presumed to be a neutron star or black hole). The
ejected mass underwent supercritical accretion on to the compact
object resulting in the subsequent ejection of much of the
material. This thereby gave rise to the observed radio emssion and the
broadened optical emission lines. Hynes et al, however, point out that
there is, so far, no evidence of binarity in this system and that,
although unlikely, the intriguing possibility exists that the
interaction with the compact object was merely a chance encounter
between two unbound objects.  CI Cam's high luminosity makes it a
Galactic counterpart to the Magellanic Could sgB[e] stars, but it does
not fit at all well into the classical Be/OB HMXB classification, and
it has even been suggested (Ishida et al 2004) that the variable soft
X-ray component might indicate that it is analogous to the supersoft
sources (chapter 11) which would make the compact object a white
dwarf.

\bigskip

{\bf Acknowledgments}\\

We are grateful to our many colleagues who commented on earlier drafts
of the manuscript and sent information in advance of publication.  In
particular, we would like to thank Reba Bandyopadhyay, Jorge Casares,
Mike Garcia, Dawn Gelino, Rob Hynes, Lydie Koch-Miramond, Tariq
Shahbaz, and John Tomsick.

\tiny

\twocolumn

\end{document}